\documentclass[a4paper]{ar2e}

\usepackage{natbib}  

\usepackage{amssymb}
\usepackage{amsmath}
\usepackage{color}
\usepackage{url}
\usepackage{graphicx}

\def\Me{{\rm M}_{\oplus}}
\def\MJ{{\rm M}_{\rm Jup}}
\def\cs{c_{\rm s}}
\def\Op{\Omega_{\rm p}}
\def\rp{r_{\rm p}}

\begin{document}

\jname{Annu. Rev. Astronomy \& Astrophysics}
\jyear{2011}
\jvol{1}
\ARinfo{1056-8700/97/0610-00}

\title{Planet-disk interaction and orbital evolution}

\markboth{Kley \& Nelson}{Planet-disk interaction}

\author{W. Kley
\affiliation{Institut f\"ur Astronomie \& Astrophysik, Universit\"at T\"ubingen,
  Auf der Morgenstelle 10, 72076 T\"ubingen, Germany}
R.P.~Nelson
\affiliation{Astronomy Unit, Queen Mary University of London, Mile End Road, London, E1 4NS, U.K.
}
}


\begin{abstract}
As planets form and grow within gaseous protoplanetary disks,
 the mutual gravitational interaction between the disk and planet
 leads to the exchange of angular momentum, and migration of the
 planet. We review current understanding of disk-planet
 interactions, focussing in particular on physical processes
 that determine the speed and direction of migration. We describe
 the evolution of low mass planets embedded in protoplanetary disks,
 and examine the influence of Lindblad and corotation torques as
 a function of the disk properties. The role of the disk in causing
 the evolution of eccentricities and inclinations is also discussed.
 We describe the rapid migration of intermediate mass planets that may
 occur as a runaway process, and examine the transition to gap formation
 and slower migration driven by the viscous evolution of the disk
 for massive planets. The roles and influence of disk self-gravity and
 magnetohydrodynamic turbulence are discussed in detail, as a
 function of the planet mass, as is the evolution of multiple planet
 systems. Finally, we address the question of how well global models
 of planetary formation that include migration are able to match 
 observations of extrasolar  planets.
\end{abstract}

\maketitle

\section{Introduction}
The discovery of numerous extrasolar planets with diverse orbital configurations
has motivated a dramatic increase in research aimed at understanding their
formation and evolution. Short-period giant planets (`hot Jupiters') such as 51 Peg b,
the first to be discovered around a Sun-like star \citep{1995Natur.378..355M}, 
are difficult to explain by {\it in situ} formation models, and 
suggest that large-scale migration has taken place.
Additional evidence for migration is provided by observations of giant planets
in mean motion resonance, such as the two giant planets orbiting the M-type star GJ~876
\citep{2001ApJ...556..296M},
or the two Saturn-mass planets in the Kepler-9 system \citep{2010Sci...330...51H}. 
Recent discoveries of resonant or near-resonant multiple systems of transiting planets 
by the Kepler mission, such as the Kepler-11 system \citep{2011Natur.470...53L}, point to an origin in a 
highly flattened and dissipative environment -- namely a protostellar/protoplanetary disk.
This review examines our current understanding of how planets interact gravitationally
with the disks within which they form.

The `nebula hypothesis' developed in the 18$^{\rm th}$ century by Kant and Laplace,
which suggests that our Solar System planets formed within a flattened and
rotating disk of gas and dust (the `solar nebula'), still forms the conceptual basis 
of modern theories of planet formation, and receives continuing support
from observations of disks around young stars. The structure and
dynamics of these disks are clearly relevant for understanding planet
formation, and we refer the interested reader to three recent reviews
that discuss the theory and observations of protoplanetary disks
\citep{2010ARA&A..48..205D, 2011ARA&A..49..195A, 2011ARA&A..49...67W}. 

There are two basic pictures of how planets form in a disk.
In a simplified version of the {\it core accretion} model, growth proceeds through a multi-stage 
process that begins with the collisional growth of submicron-sized dust grains, proceeds through 
the formation of kilometre-sized planetesimals, and leads eventually to the growth of terrestrial planets 
in the inner regions of the disk. In the outer regions, where volatiles condense
into ices, larger protoplanetary cores form that may accrete massive gaseous envelopes
to become gas giant planets. Observations suggest disk lifetimes are typically a few Myr, 
so this process must be largely completed on this timescale. The alternative {\it gravitational instability} 
model of planet formation envisages fragmentation of a protoplanetary disk into gaseous clumps
with planetary masses during the earliest stages of disk formation and evolution,
when the disk mass is comparable to that of the central star. 
Fragmentation is likely to occur at large distances in the disk ($> 50$ AU),
where the cooling time is short. In either of the above scenarios, forming planets must 
inevitably undergo gravitational interactions with their nascent disks, leading to angular momentum 
and energy exchange and orbital evolution.

Extrasolar planets may provide motivation for current research into disk-planet interactions,
but the theoretical groundwork was laid long before the
discovery of 51 Peg b \citep{1979ApJ...233..857G, 1980ApJ...241..425G, 1979MNRAS.186..799L}.
It was apparent from these early studies that low-mass planets embedded in
gaseous disks would undergo rapid inward migration (type I migration),
with Earth-mass protoplanets reaching the central star within $\sim 10^5$ yr.
Understanding how this rapid inward migration,
and the associated loss of protoplanets into the star, 
can be prevented remains an area of active research.
It has also been long recognised that giant planets 
will form gaps in their disks, and migrate inward on their
viscous evolution times (typically $10^5$ yr) \citep{1986ApJ...307..395L}.
Planetary migration is now usually divided into three categories:
{\it type I} - migration of low-mass embedded planets; {\it type II} - migration and
gap formation by massive planets; {\it type III} - rapid migration of intermediate
(Saturn) mass planets in relatively massive disks. We review each of these
migration modes, and discuss recent improvements in our understanding of
the influence of disk processes such as 
turbulence, self-gravity and thermodynamic evolution. We also discuss briefly the evolution
of orbital eccentricity and inclination driven by interaction with the disk.

Although disk-planet interactions certainly play a role in driving the
orbital evolution of forming planets, there are other mechanisms for
inducing orbital changes. Planet-planet gravitational scattering
is probably the best contender for explaining the large eccentricities of 
extrasolar planets \citep{2002MNRAS.332L..39T,2008ApJ...686..603J,2008ApJ...686..580C}, and 
the Kozai mechanism drives high eccentricities 
through secular interaction with a highly inclined companion \citep{2003ApJ...589..605W}. 
Combined with tidal interaction with the central star, these processes can
generate short-period planets on circular orbits \citep{2007ApJ...669.1298F}, and observations
suggest that these processes contribute significantly to the dynamical evolution of 
planets. Interaction with a remnant planetesimal disk, such as the Kuiper belt in our
Solar System, has also been shown to drive migration \citep{1999AJ....117.3041H,2005Natur.435..459T}.
In this review, however, we focus on the role of the gaseous disk in driving 
orbital evolution, with an emphasis on the basic physical mechanisms, recent
theoretical and computational developments, and applications to theories of
planet formation.

The important topic of planet-disk interaction has been reviewed previously by a number of researchers. 
Informative reviews that have appeared during the past few years include
\citet{2007prpl.conf..655P}, \citet{2008EAS....29..165M}, \citet{2009AREPS..37..321C},
\citet{2010exop.book..347L} and \citet{2012arXiv1203.3294B}. 
In light of these existing reviews we shall focus here on the more recent developments and 
give guidance to future issues.

\section{Planets in viscous laminar disks}
\label{sec:laminar-disks}
It is widely believed that accretion in protoplanetary disks is
driven by magnetohydrodynamical (MHD) turbulence.
However, due to the numerical complexity of solving the full time-dependent MHD equations, 
the standard means of simulating evolutionary processes in the disk is still through a simplified approach.
Here, the turbulence is modeled by assuming that its effects can be described by using the
standard Navier-Stokes equations with a Reynolds-ansatz for the viscous stress tensor. 
The viscosity parameter, often parameterized by the dimensionless number $\alpha$ \citep{1973A&A....24..337S},
is then chosen to obtain a good match with the observationally determined evolution time or
mass accretion rate of the disk. 
In this section we deal with the dynamics of embedded planets in such viscously evolving
disks. Below, in Sect.~\ref{sect:turbulence}, we focus on planets embedded in magnetised, 
turbulent disks.

\begin{figure}[ht!]
\begin{center}
 \begin{minipage}{0.52\textwidth}
 \includegraphics[width=0.99\textwidth]{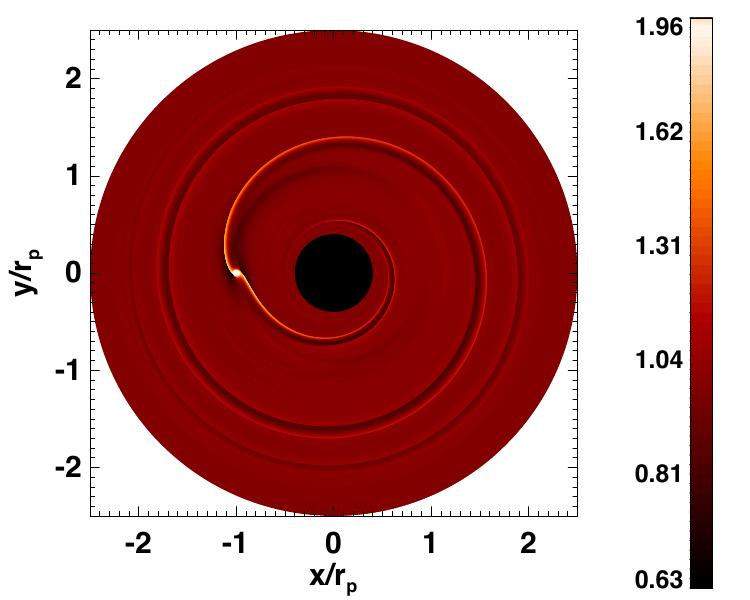}
 \end{minipage}
 $~$
 \begin{minipage}{0.45\textwidth}
 \includegraphics[width=0.99\textwidth]{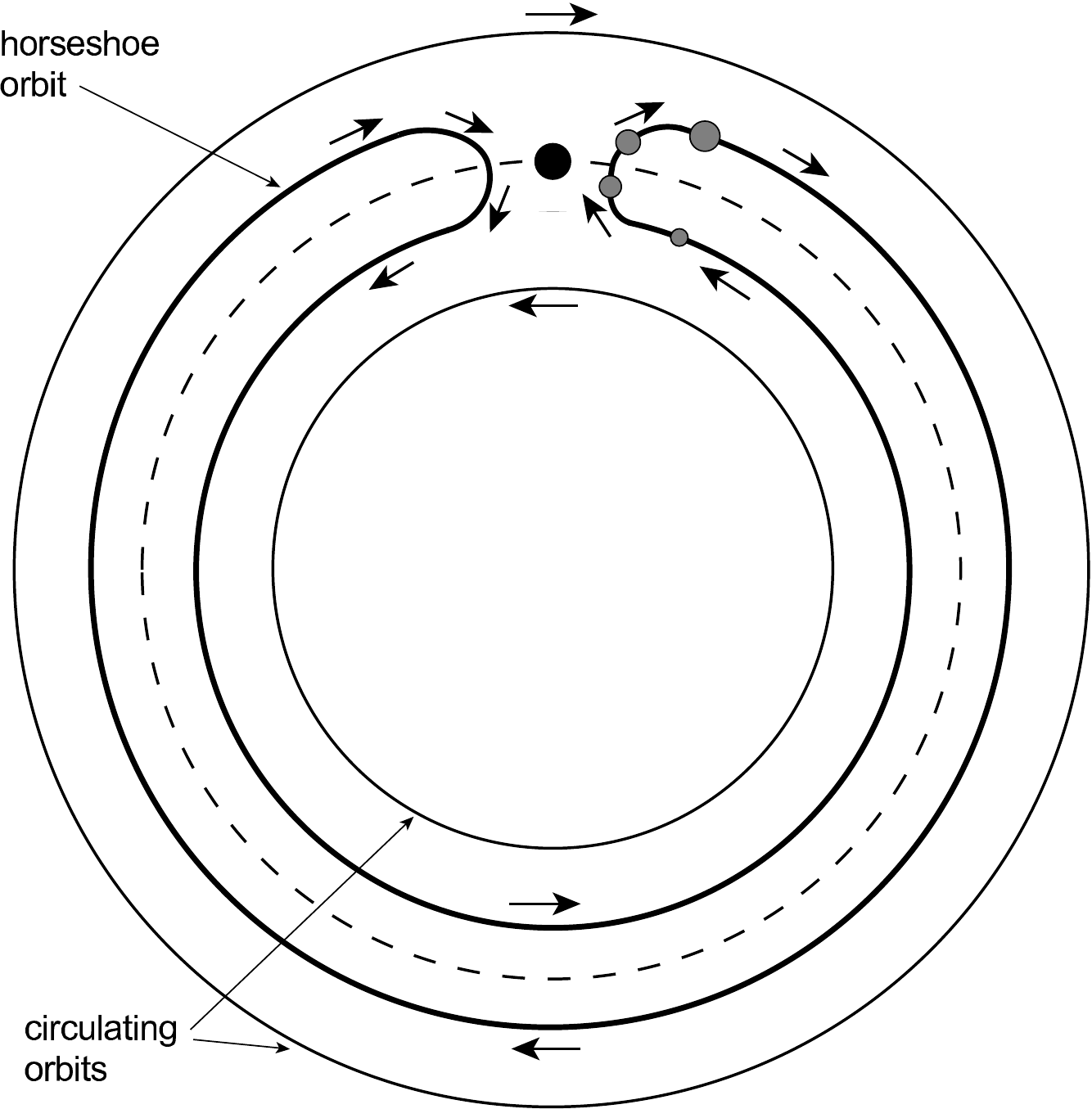}
 \end{minipage}
 \caption{Disk surface density and flow structure for a planet with mass ratio
  $q = 9 \cdot 10^{-5}$ ($30 \; \Me$ for a solar-mass star) embedded in a constant 
  surface density disk (normalised to unity).
  {\bf Left}: The density structure 5 orbits after the planet's insertion into the disk.
    Clearly visible are the spiral arms launched by the planet, the inner leading and the 
    outer trailing.
  {\bf Right}:
  Topology of the flow field, where the streamlines refer to the corotating
  frame.
  The disk is split into an inner disk with circulating streamlines (moving counterclockwise),
  a horseshoe-shaped corotation region (within the thick lines),
  and an outer disk of circulating material (moving clockwise). Courtesy P. Armitage. 
}
   \label{fig:kley-nelson-topology}
\end{center}
\end{figure}

A young planet embedded in a disk disturbs the disk dynamically in 
{\it two ways} as shown in Fig.~\ref{fig:kley-nelson-topology}:
First it divides it into an inner and outer disk separated by a {\it coorbital region} (right panel).
Second, the radially propagating density waves generated by the planet are sheared
out by the Keplerian differential rotation, creating {\it spiral waves} (left panel).
The density structure induced by the planet depends (at least for locally isothermal disks where the 
temperature scales inversely with radius)
only on the mass ratio $q = m_{\rm p}/M_*$, where $m_{\rm p}$ denotes the planet mass and $M_*$ the stellar mass.
Hence, plots like those shown in Fig.~\ref{fig:kley-nelson-topology} are often scale free and spatial units are not stated.
The perturbed density structures in the coorbital region and the spiral arms are not symmetric with respect to the line
connecting the star and the planet and gravitational torques are exerted on the planet, causing a change in its orbital elements.

Let us briefly consider the effect of the spiral arms that generate the so-called Lindblad torques
(left panel in Fig.~\ref{fig:kley-nelson-topology}). 
The inner (leading) spiral pulls the planet forward, i.e. generates a positive torque, which leads to a gain in
angular momentum of the planet and outward migration. Conversely the outer spiral pulls the planet back and drives 
inward migration. Hence, the effects of the inner and outer disk tend to partially cancel out,
and the net effect is the residual between the two sides. Because the effects of outer and inner disk are
comparable in magnitude the speed and direction of migration can depend delicately on the physical details of the disk.
The torque on the planet can be calculated approximately by looking at the angular momentum exchange of individual particles
passing by the planet: This is the so-called impulse approximation [for details see \citet{1979MNRAS.186..799L} and \citet{2010exop.book..347L}].
However, a full fluid dynamical approach leads to deeper insight and to more reliable results. We follow this latter approach and 
present an outline of how the torques are evaluated below.

\subsection{Low-mass planets in isothermal disks}
For {\it low-mass embedded planets} that induce only small perturbations in the disk,
the angular momentum exchange between disk and planet can be determined
from a {\it linear analysis} of the perturbed flow. 
The unperturbed disk is assumed to be axisymmetric, and in a state of
Keplerian rotation with angular velocity $\Omega(r) = \sqrt{G M_*/r^3}$ around the
star, and with vanishing radial velocity.
For now, we consider a planet on a {\it circular orbit}. Its
gravitational potential, $\psi_{\rm p}$, is periodic in azimuth
and can be expanded in a Fourier series:
\begin{equation}
\label{eq:kley-nelson-psi}
    \psi_{\rm p} (r,\varphi,t) \equiv  - \frac{G m_{\rm p}}{|\vec{r}_{\rm p}(t) - \vec{r}|} \, = \, 
      \sum_{m=0}^\infty\psi_m(r)\cos\{m[\varphi-\varphi_{\rm p}(t)]\},
\end{equation}
where $\varphi_{\rm p}=\Op t$ is the azimuth angle of the planet, moving with
angular velocity $\Op$. $\psi_m(r)$ denotes the potential coefficient for each
azimuthal mode number $m$. Each potential component rotates with pattern-speed
$\Op$. An explicit expression for $\psi_m(r)$ is given, for example, by \citet{1987Icar...69..157M}.
The above decomposition applies to a flat two-dimensional disk and can
be generalized for a planet on an eccentric orbit (see Sect.~\ref{subsect:orb-elem} below).
The total torque exerted by the disk {\it on the planet} can 
be calculated according to
\begin{equation}
\label{eq:gamma_tot}
      \Gamma_{\rm tot} = - \int_{\rm disk} \Sigma ( \vec{r} \times \vec{F}) \, df  \, = \,
      \, \int_{\rm disk} \Sigma ( \vec{r} \times \nabla \psi_{\rm p}) \, df  \, = \,
      \, \int_{\rm disk}  \Sigma \frac{\partial \psi_{\rm p}}{\partial \varphi} df \,,
\end{equation}
where $\Sigma$ denotes the surface density of the disk,
$\vec{F}$ is the specific force (acceleration) acting on a disk element, and $df$ is a surface element. 
Whenever the frequency of an individual potential component as seen by a
fluid particle in the disk,
$\omega=m(\Omega(r)-\Op)$, matches a natural oscillation frequency,
we have a resonant condition inducing a strong disk response. 
Torques are therefore calculated at these resonant locations.

Neglecting pressure and self-gravity, resonances occur for
$m(\Omega(r)-\Op) = 0$ or $\pm \kappa(r)$, where $\kappa(r)$ is 
the epicyclic frequency in the disk, i.e. the oscillation frequency for a particle in the
disk subject to a small radial displacement. The first case refers to the
corotation resonance where the local disk and planet orbital speeds are equal, 
$\Omega(r)=\Op$. 
The corotation region is indicated in the right panel of 
Fig.~\ref{fig:kley-nelson-topology} within the thick line.
The second resonance condition applies to {Lindblad resonances} where 
the plus sign, i.e. $\Omega(r) = \Op + \kappa(r)/m$, refers to {\it inner} Lindblad resonances 
(interior to $r_{\rm p}$), where the disk rotates faster than the planet. The minus sign
refers to the {\it outer} Lindblad resonances, i.e. $\Omega(r) = \Op - \kappa(r)/m$ 
(exterior to $r_{\rm p}$ and the corotation region in Fig.~\ref{fig:kley-nelson-topology}). 
The radial locations of the Lindblad resonances, $r_{\rm L}$, are obtained by noting that for Keplerian 
disks, $\kappa = \Omega$, which leads to
\begin{equation}
                r_{\rm L} =  \left( \frac{m}{m \pm 1} \right)^{2/3} \, r_{\rm p}.
\label{eq:LR-positions1}
\end{equation}
Under conditions where pressure in a disk can not be neglected,
the Lindblad resonance condition is modified to become
\begin{equation}
\label{eq:reson-press}
   m(\Omega(r) - \Op) =\sqrt{\kappa^2(r)(1 + \xi^2)}.
\end{equation}
Here, $\xi \equiv m \cs/(\Omega r)$ where $\cs$ is the isothermal sound speed in the disk.
For $m \rightarrow \infty$, and denoting $\cs = H \Omega$ where
$H$ is the local disk scale height (see eq.~\ref{eq:kley-nelson-thickness} below), 
the Lindblad resonance positions are shifted away from those given by 
eq.~(\ref{eq:LR-positions1}) to become
\begin{equation}
r_{\rm L} = r_{\rm p} +\frac{2 H}{3},
\label{eq:LR-positions2}
\end{equation}
showing that the Lindblad resonances for $m \gg 1$ pile-up at
a distance equal to $2H/3$ from the planet. This prevents 
divergence of the torque experienced by the planet from the disk,
giving rise to the phenomenon known as the {\it torque cut-off}
\citep{1980ApJ...241..425G,1993ApJ...419..155A}.

\subsubsection{Lindblad torques}
Spiral density waves are excited by the planet at Lindblad resonances,
and carry an angular momentum flux as they propagate away from the
resonance into the disk. 
The combined effect of both spirals determines the sign and magnitude of 
the total torque. 
For circular orbits the total torque exerted on the planet is a
direct measure of the speed and direction of migration, with positive torques
driving outward migration and negative ones driving migration inward.

The Lindblad torque acting {\it on the planet} due to the disk response to
the $m^{\rm th}$ component of the planet potential is
\begin{equation}
   \Gamma^{\rm L}_m = {\rm sign}(\Omega-\Op) \,  \frac{\pi^2 \Sigma}{3 \Omega \Op} \, 
   \left(r \frac{d \psi_m}{dr} + \frac{2 m^2 (\Omega - \Op)}{\Omega} \, \psi_m 
   \right)^2\,,
\label{eq:gamma_L}
\end{equation}
which has to be evaluated at the resonance location $r=r_{\rm L}$.
Details of the derivation of this, or equivalent expressions, can be found for example
in \citet{1979ApJ...233..857G} \cite{1986Icar...67..164W} and \cite{1987Icar...69..157M}.
In principle a correction factor of $\left[ \sqrt{1 + \xi^2} (1 + 4 \xi^2) \right]^{-1}$
should be applied to eq.~(\ref{eq:gamma_L}) to account for the shift in resonances
described by eq.~(\ref{eq:reson-press}) \citep{1993ApJ...419..155A,1997Icar..126..261W}.
Equation (\ref{eq:gamma_L}) is for a single Fourier component
of $\psi_{\rm p}$ acting at an inner or outer Lindblad resonance. The total
torque acting on the planet, usually referred to as the differential
Lindblad torque, is then a sum over contributions
for $1 \le m \le \infty$ at the inner and outer Lindblad resonances.
For most disk models the differential Lindblad torque is negative,
driving inward migration of the planet because the outer Lindblad
resonances lie closer to the planet than the inner resonances.
In particular, gas pressure in the disk causes the angular velocity to
be slightly sub-Keplerian, shifting the locations of both inner and outer 
Lindblad resonances inward \citep{1997Icar..126..261W,2007prpl.conf..655P}.

We comment briefly that the influence of disk self--gravity on type~I migration 
of low-mass embedded planets has been considered by \cite{2005A&A...433L..37P}
and \cite{2008ApJ...678..483B}. Their studies indicate that a
modest shift of Lindblad resonance locations arises when both the disk
and planet orbit self-consistently in the gravitational field of
the star and disk, leading to a slowing of type~I migration by a 
factor of $\sim 2$ in a disk with mass equal to 3 times the minimum mass
solar nebula model (MMSN) \citep{1981PThPS..70...35H}.

\subsubsection{Corotation torques}
Material within the horseshoe region 
(inside the thick line in Fig.\ref{fig:kley-nelson-topology}) has (on average) the same speed as the
planet and is corotating with it. In the limit of small planet masses (a few $\Me$ at 5~AU) one can apply
linear theory and calculate the corresponding corotation torque.
This was first estimated in the context of planet-satellite interaction by
\citet{1979ApJ...233..857G} with the result that
\begin{equation}
\label{eq:gamma_C}
    \Gamma^{\rm C}_m = 
      \frac{m \pi^2}{2} \, 
      \frac{\psi_m}{r d\Omega/dr} \, 
      \frac{d}{dr} \left( \frac{\Sigma}{B} \right),
\end{equation}
which has to be evaluated at the corotation radius, $r=r_{\rm C}$. Here, $B=\kappa^2/(4 \Omega)$ 
denotes the second Oort constant.
Physically, $B$ is half the $z$-component of the flow vorticity, 
$(\nabla \times \vec{v})|_z$ (i.e. $2 B/\Sigma$ is the
specific vorticity, sometimes called the vortensity).

\subsubsection{Type~I migration in isothermal disks}
Low-mass planets do not alter the global disk structure significantly, and in particular they do not open gaps.
Hence, the combined effect of Lindblad and corotation torques can be calculated for small planetary masses using
the above {\it linear analysis} (we discuss the validity of this approach below). 
The outcome of such linear studies has been termed {\it type~I migration}.
Due to the complexity of considering heat generation and transport in disks these linear studies have
relied nearly exclusively on simplified, isothermal disk models, where the disk has a fixed radial temperature structure,
$T(r)$. If $dT/dr \neq 0$, then the models have been termed {\it locally isothermal},
where the temperature varies radially but is constant
in the vertical direction if one considers the three-dimensional (3D) structure.
Vertical hydrostatic equilibrium leads to a Gaussian density stratification.
Linear calculations have been performed for both 2D (flat disks using $r\varphi$-coordinates)
and full 3D configurations, but in 2D one must
account for the vertical distribution of disk matter by using an effective smoothing of the gravitational
potential near the planet (aside from the necessity when simulating disk-planet interaction).

Comprehensive 3D linear calculations have been presented by \citet{2002ApJ...565.1257T}, and
for {\it strictly isothermal} disks with $T = {\rm constant}$,
they yield the following expressions for the differential Lindblad and corotation 
torques acting {\it on} the planet:
\begin{equation}
\label{eq:kley-nelson-gammatot}
    \Gamma^{\rm L}_{\rm lin}  =  - (2.34 - 0.1 \beta_\Sigma) \, \Gamma_0   
      \quad  \quad \mbox{and} \quad \quad 
    \Gamma^{\rm C}_{\rm lin}  =  0.64 \, \left( \frac{3}{2} - \beta_\Sigma\right) \Gamma_0.
\end{equation}
Here, the surface density varies with radius as $\Sigma(r) \propto r^{-\beta_\Sigma}$, and
the torque normalization is given by
\begin{equation}
\label{eq:kley-nelson-gamma0}
       \Gamma_0 =  \left(\frac{m_{\rm p}}{M_*}\right)^{2}
       \, \left(\frac{H}{r_{\rm p}}\right)^{-2}   \, \Sigma_{\rm p} \, \rp^4 \, \Op^2,
\end{equation}
which must be evaluated at the planet location, as indicated by the index $p$.
The total torque, $\Gamma_{\rm lin}^{\rm tot}$, is given as the sum of the
differential Lindblad and corotation torques,
$\Gamma_{\rm lin}^{\rm tot} = \Gamma_{\rm lin}^{\rm L} + \Gamma_{\rm lin}^{\rm C}$ \citep{1986Icar...67..164W}. 
The magnitude of the type~I torque scales as the inverse square of the disk aspect
ratio (or inversely with the temperature, see eq.~\ref{eq:kley-nelson-thickness}),
quadratically with the planet mass, and linearly with the disk mass. The quadratic 
$m_{\rm p}$ dependence arises because density perturbations scale with $m_{\rm p}$ 
and another factor of $m_{\rm p}$ comes in to play when evaluating the force or torque on the planet.

We define the effective disk thickness, $H$, in this review as
\begin{equation}
\label{eq:kley-nelson-thickness}
        H(r) = c_{\rm s} \Omega^{-1},
\end{equation}
where $c_{\rm s}^2 = P/\Sigma$ is the local {\it isothermal} sound speed, and $P$ is the vertically integrated pressure
for 2D disk models. 
For adiabatic disks the sound speed is given by $\sqrt{\gamma} \, c_{\rm s}$ with
adiabatic exponent $\gamma$. 
An interesting feature of the corotation torque is that it vanishes for the
popular MMSN protoplanetary disk model. The surface
density in the MMSN scales as $\Sigma \propto r^{-3/2}$ giving $\Gamma^{\rm C} = 0$, such that only the
differential Lindblad torque remains.

\subsubsection{Numerical simulations of type~I migration}
The linear studies have been supplemented in recent years by many numerical studies in two and three dimensions.
Here, the disk is modeled as a viscous gas and the full non-linear Navier-Stokes equations are solved,
typically (but not exclusively) using grid-based hydrodynamic codes.
Because there are a number of codes on the market, a detailed code comparison
project was conducted a few years ago to verify the convergence of the results 
\citep{2006MNRAS.370..529D}. Because the physical size of an embedded planet can be much smaller than
the achievable grid resolution, it is typically treated as a point mass. In this case, the
gravitational potential has to be smoothed (over a couple of grid-cells) to avoid singularities.
Mostly, the following form is used:
\begin{equation}
\label{eq:pot-smooth}
     \psi_{\rm p} = - \, \frac{G M_* m_{\rm p}}{(s^2+\epsilon^2)^{1/2}},
\end{equation}
where $s$ is the distance from the planet, and $\epsilon$ the smoothing length.
As mentioned above, in 2D the neglect of the vertical dimension
requires a larger value of $\epsilon \approx H/2$.

Numerical improvements have led to increased computational efficiency and accuracy that allow longer and
higher resolution simulations to be performed. Worthy of mention are the {\tt FARGO} method
\citep{2000A&AS..141..165M}, which overcomes restrictions on the time-step due to
the differentially rotating disk, and a conservative treatment of the Coriolis force \citep{1998A&A...338L..37K}.
To increase spatial resolution around the planet,
nested-grid structures have been employed successfully in 2D and 3D 
\citep{2002A&A...385..647D,2003ApJ...586..540D}.
The most recent 3D nested-grid hydrodynamic simulations of planet-disk interaction
have been presented by \citet{2010ApJ...724..730D} for {\it locally isothermal} disks.
Their simulations give very good agreement with the 3D linear results,
and the range of
applicability of eq.~(\ref{eq:kley-nelson-gammatot}) is extended by considering the influence
of radial temperature gradients in the disk, such that $T(r)=T_0 \, r^{-\beta_T}$.
Their calculations yield the following form for the total (Lindblad and corotation) torque acting
on planets with masses below about 10 $\Me$:
\begin{equation}
\label{eq:kley-nelson-gennaro}
        \Gamma_{\rm tot}  = - (1.36 + 0.62 \beta_\Sigma + 0.43 \beta_T)  \Gamma_0.
\end{equation}


The torques acting on the planet change its angular momentum $J_{\rm p}$ according to
\begin{equation}
\label{eq:kley-nelson-lp}
         \frac{d J_{\rm p}}{d t} = \Gamma_{\rm tot}.
\end{equation}
For circular orbits, $J_{\rm p}$ depends only on the planet's distance $a_{\rm p}$ (semi-major axis)
from the star through $J_{\rm p} = m_{\rm p} \sqrt{G M_* a_{\rm p}}$.
Equation~(\ref{eq:kley-nelson-lp}) can then be solved for the rate of change in the 
semi-major axis, and one obtains the migration rate $\dot{a}_{\rm p}$ of the planet.
The migration timescale, $\tau_{\rm mig}$, is then
\begin{equation}
\label{eq:kley-nelson-taumig}
       \tau_{\rm mig} =  \frac{a_{\rm p}}{\dot{a}_{\rm p}} = \frac{1}{2} \frac{J_{\rm p}}{\Gamma_{\rm tot}}.
\end{equation}

In early studies \citep{1980ApJ...241..425G} realised that the
migration timescale can be very short.
For an Earth-mass planet around a solar-mass
star at 1~AU, we find for the MMSN ($\Sigma = 1700$ g cm$^{-3}$, $H/r=0.05$) 
$\tau_{\rm mig} \approx 10^5$ yr. For planetary cores of a few Earth-masses initially
located at 5 AU, the migration timescale is shorter than the gas accretion time onto the core,
which is typically a few Myr \citep{1996Icar..124...62P}. This realisation has
posed a serious challenge for the theory of planet formation over recent decades, as we discuss in
more detail in Sect.~\ref{sect:migrat-obs}. Possible remedies that have been suggested for the problem
of excessively rapid type~I migration include: strong corotation torques operating
in regions where there are steep positive surface density gradients (so-called `planet traps')
\citep{2006ApJ...642..478M}; reductions in the differential Lindblad torque in regions of
sharp opacity transition \citep{2004ApJ...606..520M}; magnetic resonances in disks with
strong toroidal magnetic fields \citep{2003MNRAS.341.1157T}; 
torque reversals for planets on eccentric orbits \citep{2000MNRAS.315..823P,2006A&A...450..833C};
torque reductions from disk shadowing and illumination variations in the presence 
of a planet \citep{2005ApJ...619.1123J}; stochastic migration induced by disk turbulence
\citep{2004MNRAS.350..849N, 2004ApJ...608..489L}; and corotation torques in radiative disks
\citep{2006A&A...459L..17P}. We discuss some of these ideas in greater detail in
the following sections.

\begin{figure}[ht!]
\begin{center}
 \begin{minipage}{0.49\textwidth}
 \includegraphics[width=0.95\textwidth]{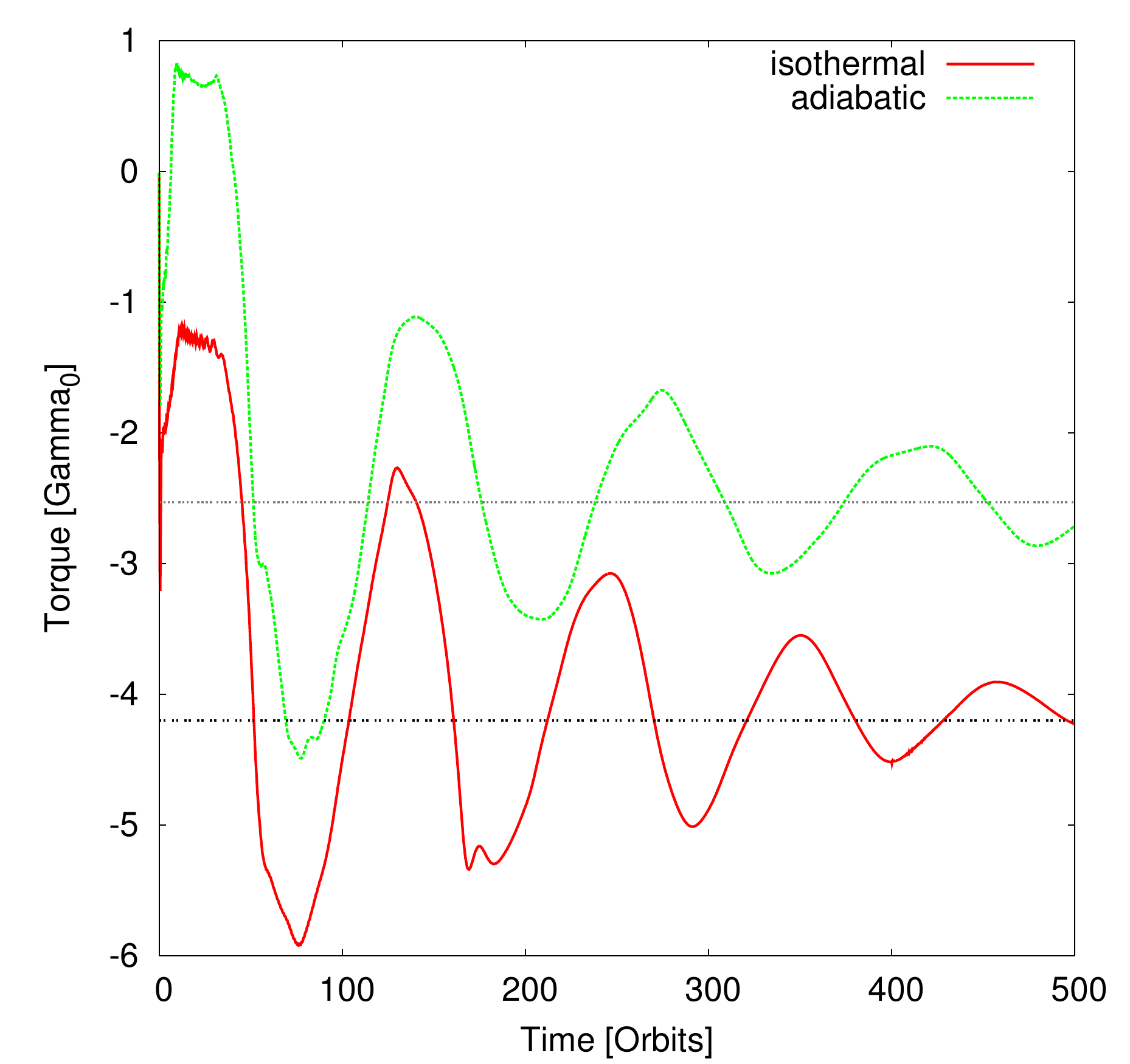}
 \end{minipage}
 \begin{minipage}{0.49\textwidth}
 \includegraphics[width=0.99\textwidth]{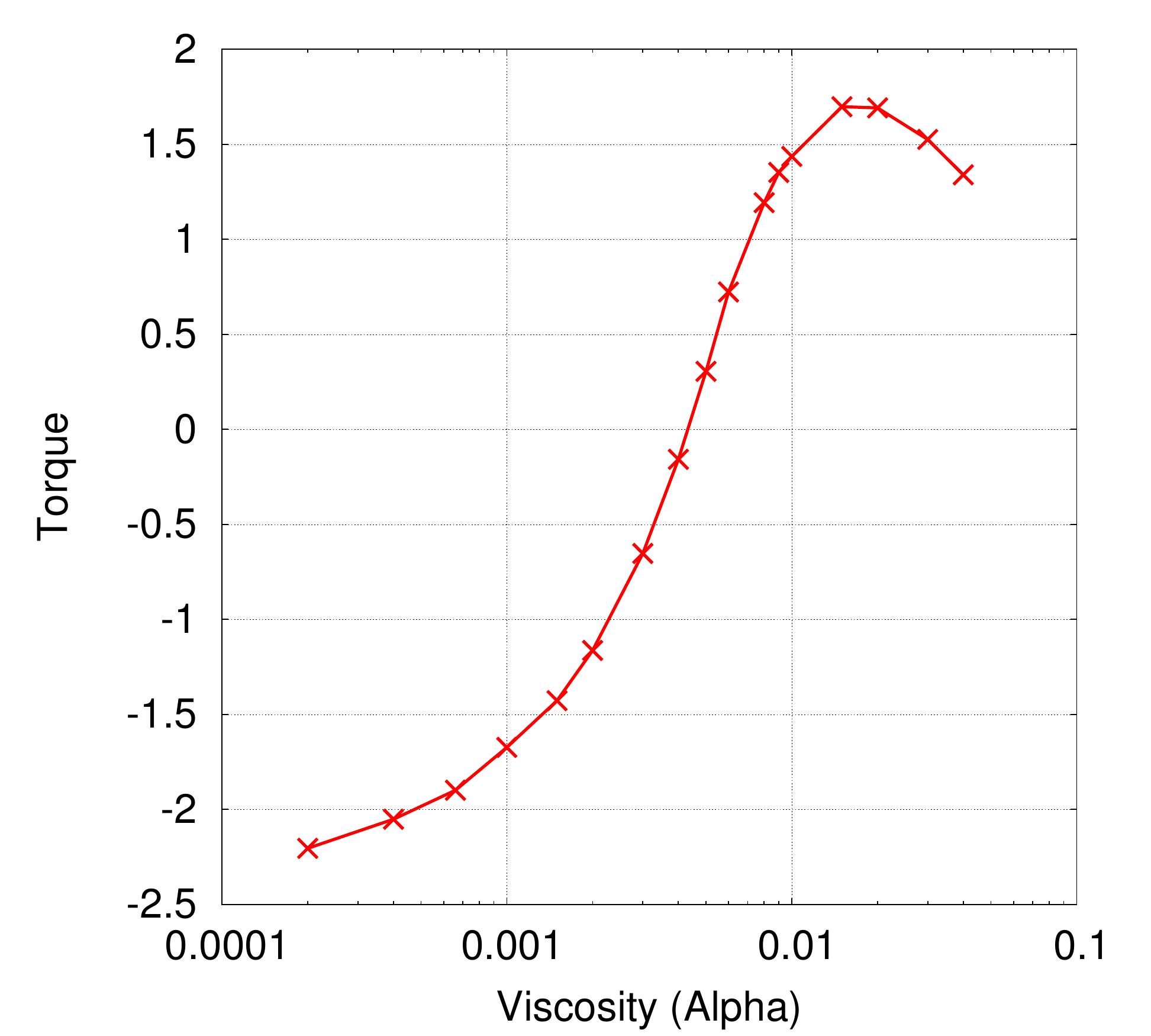}
 \end{minipage}
 \caption{Torque saturation:
  {\bf Left:} 
   Time evolution of the total torque acting on a low mass planet
  in a disk that has a constant surface density and very low viscosity.
  Shown are results for an adiabatic and a locally isothermal disk.
  At the start of the simulations, when the gas performs its first U-turn near the planet,
  the torques reach a maximum (fully unsaturated values).
  Then they drop to reach the final saturated values through a series of oscillations.
  The horizontal lines refer to the isothermal and adiabatic Lindblad torques
  \citep{2010MNRAS.401.1950P}.
  {\bf Right:} 
   Final equilibrium torque acting on a 20 $\Me$ planet
  embedded in a viscous disk for 2D radiative simulations.
  The viscosity is given by $\nu = \alpha c_{\rm s} H$ with $\alpha= 4 \cdot 10^{-3}$.
  For small viscosity the torque is fully saturated and given by the 
  Lindblad torque. Increasing the viscosity desaturates the corotation torque
  such that the total torque increases, whereas at very large $\alpha$ the viscosity perturbs 
  the horseshoe region and the torque declines again.
 }
   \label{fig:kley-nelson-saturat}
\end{center}
\end{figure}

\subsubsection{The horseshoe drag}
\label{subsubsec:horse-drag}
In the discussion of the corotation torque above, it was assumed that it can be estimated by linear
analysis.
This approach has been questioned recently by \citet{2008A&A...485..877P}, who point out that the corotation torque is
non-linear for all planet masses. The argument is based on an analysis of the so-called horseshoe drag 
\citep{1991LPI....22.1463W}, which is a description of the corotation torque based on
horseshoe orbits (that do not exist in a linear theory).
The right panel of Fig.~\ref{fig:kley-nelson-topology} shows the motion of the gas in the frame
corotating with the planet.
Here, fluid elements approach the planet at the two ends of the horseshoe orbit and perform U-turns. 
They are periodically shifted from an orbit with a semi-major axis slightly
larger than the planet's to an orbit that is slightly smaller, and vice versa. Hence, at each U-turn 
there is an exchange of angular momentum between coorbital disk material and the planet.
The total corotation torque is obtained by adding the contributions from each of the horseshoe U-turns.
In the case of pressureless particles that do not interact, there is no net torque exerted
on the planet because the particles follow the trajectories of the restricted three-body problem.
To obtain a non-zero net torque, a persistent asymmetry between the two U-turns must exist. 
The strength of the asymmetry in the horseshoe region depends on the radial gradients
of specific vortensity and entropy across it
\citep{1991LPI....22.1463W,2008ApJ...672.1054B,2008A&A...485..877P}.

The expression for the horseshoe drag obtained by \citet{1991LPI....22.1463W} is  
\begin{equation}
\label{eq:kley-nelson-gamma-hs}
    \Gamma^{\rm HS}  = \frac{3}{4} \left( \frac{3}{2} - \beta_\Sigma  \right) \, x_{\rm s}^4 
 \, \Sigma_{\rm p} \, \Op^2.
\end{equation}
The sensitivity to the geometry of the corotation region is indicated by the scaling of $\Gamma^{\rm HS}$ with
the 4$^{\rm th}$ power of the radial half-width of the horeshoe region, $x_{\rm s}$ (Fig.~\ref{fig:kley-nelson-topology} right panel).
Comparing this to the expression for $\Gamma^{\rm C}_{\rm lin}$ in eq.~(\ref{eq:kley-nelson-gammatot}),
we can see immediately that the two expressions are not identical.
The horseshoe half-width, $x_{\rm s}$, is normally determined through an analysis of
fluid streamlines in numerical simulations  \citep{2002A&A...387..605M}.
A number of numerical studies \citep{2006ApJ...652..730M,2008ApJ...672.1054B,2008A&A...485..877P}
indicate that an appropriate form for $x_{\rm s}$ (for 2D disks) is given by
\begin{equation}
\label{eq:kley-nelson-xs}
     x_{\rm s}  = C(\epsilon) \, a_{\rm p} \, \sqrt{ \frac{q}{h} },
\end{equation}
where $C(\epsilon)$ is a factor of order unity and $\epsilon$ is the smoothing length given in eq.~(\ref{eq:pot-smooth}).
We note that the relative scale height $h=H/r$ depends on the
sound speed $c_{\rm s}$ (see eq.~\ref{eq:kley-nelson-thickness}), and hence the value of $x_{\rm s}$ in
adiabatic disks is smaller than in isothermal disks by a factor of $\gamma^{1/4}$.
Using eq.~(\ref{eq:kley-nelson-xs}) in eq.~(\ref{eq:kley-nelson-gammatot}) and comparing
with eq.~(\ref{eq:kley-nelson-gamma-hs})
shows that $\Gamma^{\rm C}_{\rm lin}$ and $\Gamma^{\rm HS}$ scale identically with physical parameters, and differ
only by a constant factor. In general, the horseshoe drag is larger than the linear corotation
torque, and therefore tends to be more effective in slowing down, or even reversing the inward migration of
low mass planets.

\subsubsection{Corotation torque saturation}
\label{subsubsec:corot-sat}
As mentioned above, the corotation torque depends on radial gradients of specific vorticity and
entropy across the corotation region.
If viscosity or thermal diffusion are too small, the corresponding contributions
to the corotation torque will saturate \citep{1992NYASA.675..314W},
which means that the torque strength can be greatly reduced or even vanish completely.
Fluid elements at different radii within the corotating region have different horseshoe libration periods, 
and this leads to phase mixing and flattening of the original specific vorticity or entropy
gradients. Viscosity acts to re-establish the specific vorticity gradient, 
so an appropriate level of viscosity acts to desaturate the vorticity-related 
corotation torque.  Heat diffusion acts to re-establish the entropy gradient,
so an appropriate level desaturates the entropy-related corotation torque.
The horseshoe region contains only a finite amount of angular momentum
that can be exchanged with the planet in the absence of viscosity \citep{2001MNRAS.326..833B}.
Therefore viscosity is always required to desaturate the corotation torque
even if the vorticity-related torque is inactive (as it would be in a disk where 
$\Sigma = \Sigma_0 r^{-3/2}$, for example). In this case the role of viscosity is to exchange angular 
momentum between the horseshoe region and the rest of the disk.

The saturation process is shown in Fig.~\ref{fig:kley-nelson-saturat} (left panel) where the torque evolution for a 4.2-$\Me$
planet embedded in a constant surface density disk with very low disk viscosity (here $\nu = 10^{-8} a_{\rm p}^2 \Op$)
is shown for an isothermal and adiabatic simulation.
Such a disk has a specific vorticity profile that scales as $r^{-3/2}$, so we expect a strong vorticity-related
torque.
Indeed, shortly after the start of the simulations, the gas close to the planet begins to follow horseshoe orbits and generates
a strong positive horseshoe drag as described above. This is because the initial gradient in
specific vorticity generates large asymmetries in the torques exerted at the two horseshoe
U-turns, which are fully established after about 10-20 orbits.
Later, after the material has completed approximately half a horseshoe libration, phase mixing begins to set in and
the torque drops strongly.  After a few oscillations the torque approaches a stationary state. 
The oscillations are due to the consecutive mixing of the coorbital material within the horseshoe region,
and their period is roughly equal to the libration time of the material near the outer
edges of the horseshoe region, which can be estimated by 
\begin{equation}
\label{eq:kley-nelson-tlib}
       \tau_{\rm lib} = \frac{8 \pi \, a_{\rm p}}{3 \, \Op x_{\rm s}}.
\end{equation}
In our case (with parameters $h=0.05$, $q=1.26 \times 10^{-5}$, and $C=1$ in eq.~\ref{eq:kley-nelson-xs}),
one finds an oscillation timescale of $\sim 85$ orbits for the isothermal and $\sim 100$ orbits for the adiabatic case,
in good agreement with the simulations.
In the final state, the corotation torque/horseshoe drag are fully saturated
and only the differential Lindblad torque remains.
As indicated by the torque expressions given above in 
eqs.~(\ref{eq:kley-nelson-gammatot}) and (\ref{eq:kley-nelson-gamma0}),
the isothermal Lindblad torque is a factor $\gamma$ stronger than the adiabatic one.
The physical reason
lies in the larger adiabatic sound speed (i.e. larger $H$), which causes wider spiral arms that are launched further from the 
planet and, hence, weaken the torques. 

Desaturation of the corotation torque requires the action of viscosity,
and the optimal level of desaturation occurs when the viscous diffusion time scale across
the horseshoe region, $\tau_{\nu} \lesssim \tau_{\rm lib}$ 
\citep{2001ApJ...558..453M,2002A&A...387..605M}, where $\tau_{\nu} = x_{\rm s}^2/\nu$,
and $\nu$ is the kinematic viscosity. Similar arguments apply to the desaturation
of the entropy-related corotation torque, but the controlling parameter in that
case is the thermal diffusion coefficient.
The dependence of desaturation on the viscosity is shown in Fig.~\ref{fig:kley-nelson-saturat} (right panel),
where it can be seen that the torque decreases (i.e. saturates) as the viscosity gets smaller.
In very viscous disks, when the viscous diffusion time becomes short and comparable to the time 
for the gas to undergo a horseshoe U-turn, the non-linear corotation torque starts to
weaken (Fig.~\ref{fig:kley-nelson-saturat}) as the viscosity effectively disrupts the horseshoe streamlines.
Increasing the viscosity further causes the vorticity-related corotation torque to approach
the smaller value expected from linear theory \citep{2002A&A...387..605M}.
A similar argument applies to the saturation of the entropy-related 
corotation torque when thermal diffusion becomes increasingly efficient 
\citep{2008A&A...485..877P,2011MNRAS.410..293P}. 

\subsection{Migration in adiabatic and radiative disks}
\label{subsec:adi-rad}
As mentioned above, most previous studies of disk-planet interaction have been performed in
the context of (locally) isothermal disks. However, as shown in
Fig.~\ref{fig:kley-nelson-saturat}, a process that prevents corotation
torque saturation in an adiabatic disk could change migration significantly, possibly
even reversing it.
Indeed, in a seminal study, \citet{2006A&A...459L..17P} have demonstrated exactly this possibility 
when considering planets embedded in 3D disks including radiation transport.
Inspired by this work, a number of studies on planet-disk interaction have been devoted to examining
migration in more realistic, non-isothermal disks. Through a linear analysis and accompanying 2D
simulations, \citet{2008ApJ...672.1054B} have shown that in adiabatic disks the (unsaturated) corotation
torque scales with the radial gradient of the entropy. This comes about because in adiabatic flows the entropy, 
$S = P/\Sigma^{\gamma}$, is conserved along streamlines. Hence, upon executing horseshoe U-turns, an underlying
entropy gradient results in a density disturbance because the disk tries to maintain
local pressure equilibrium. As density varies inversely
with entropy, the torque is proportional to the negative radial gradient of the entropy. 
This effect can lead to positive unsaturated torques, as shown by the early torque evolution in
Fig.~\ref{fig:kley-nelson-saturat} (green curve in left panel).
However, due to phase mixing in the horseshoe region the torque saturates because of
the vanishing entropy gradient across it. Thermal diffusion is required to maintain the entropy gradient 
\citep{2008ApJ...672.1054B,2008A&A...485..877P}, as described above.

To demonstrate the effect for more realistic 2D flat disks, we show results for a planet-disk simulation where
we include viscous heating, and local radiative cooling as well as grey diffusive radiative transport in the disk midplane
(we refer to models that include radiative cooling and/or radiation transport as radiative disks).
The energy equation reads
\begin{equation}
\label{eq:kley-nelson-energy}
 \frac{\partial \Sigma c_{\rm v} T}{\partial t} + \nabla \cdot (\Sigma c_{\rm v} T {\bf u} )
       =  \,  - P  \nabla \cdot {\bf u} 
       \, + D  - Q 
      \,  - 2 H \nabla \cdot \vec{F} 
\end{equation}
where $T$ is the midplane temperature, $D$ the viscous dissipation,
$Q$ the radiative cooling, and $\vec{F}$ the radiative flux in the midplane. Models where the various contributions
on the right hand side of eq.~(\ref{eq:kley-nelson-energy}) were selectively switched off and on, have been presented
in \citet{2008A&A...487L...9K}.
\begin{figure}[ht!]
\begin{center}
\begin{minipage}{0.50\textwidth}
 \includegraphics[width=0.99\textwidth]{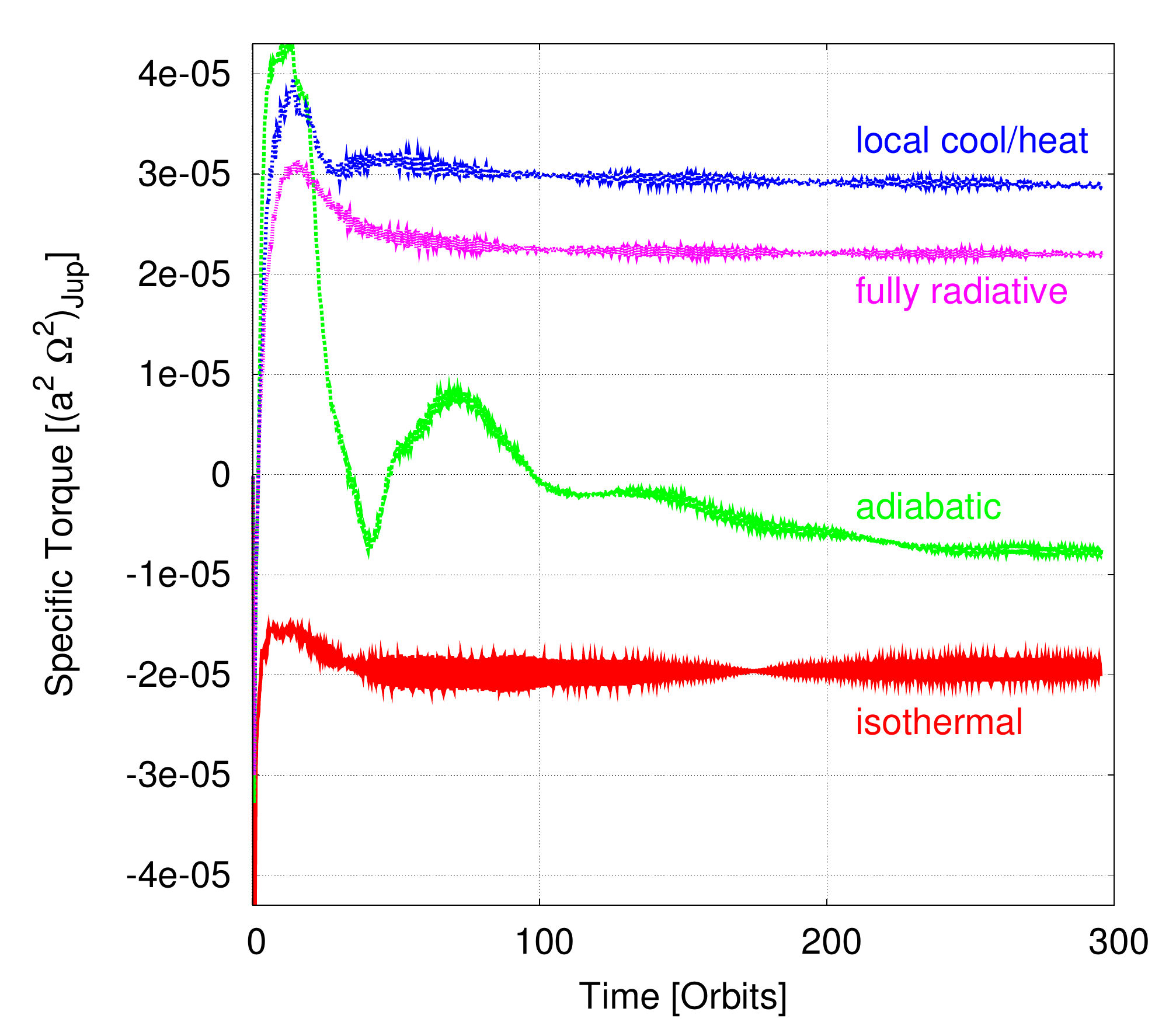}
\end{minipage}
\begin{minipage}{0.49\textwidth}
 \includegraphics[width=0.99\textwidth]{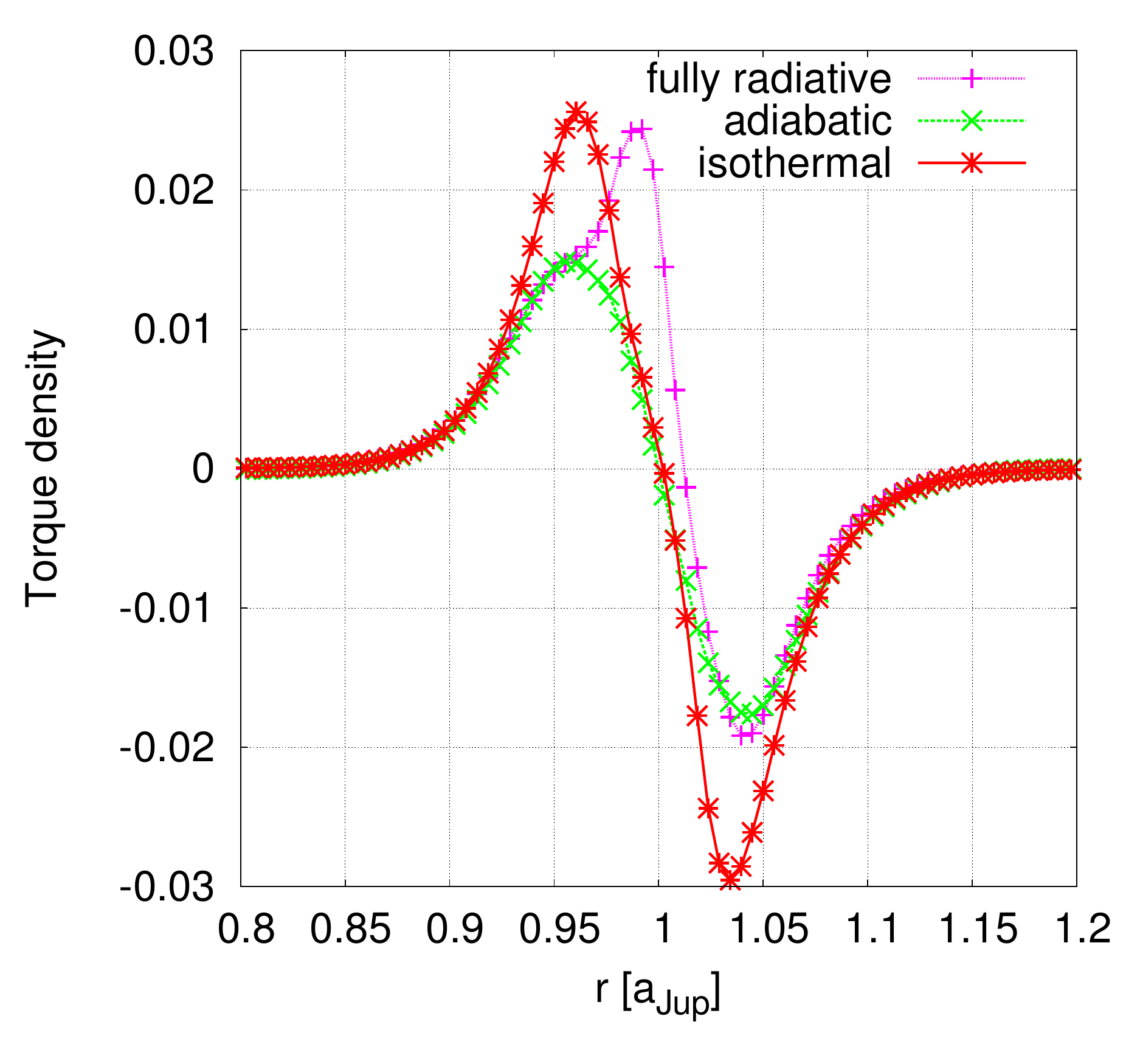}
\end{minipage}
 \caption{
  Torque acting on a 
  a 20 $\Me$ planet at $5.2$AU in a disk with 0.01 solar masses
 within $2.08$ and $13$ AU.
  {\bf Left:}
  Time evolution of the specific torque (per planet mass) for 2D simulations.
  Shown are 4 cases with a different treatment of the gas thermodynamics, see eq.~(\ref{eq:kley-nelson-energy}):
  {\it i}) no energy equation (isothermal),
  {\it ii}) with only the first term on the rhs. of eq.~(\ref{eq:kley-nelson-energy}) (adiabatic),
  {\it iii}) with all terms on the rhs. (fully radiative), and
  {\it iv}) with all but the last term on the rhs. (local cooling/heating), \citep[after][]{2008A&A...487L...9K}.
  {\bf Right:}  
  Radial torque density $d \Gamma(r)/dm$ 
  for full 3D simulations, scaled in units of $(d \Gamma/dm)_0$ (eq.~\ref{eq:gamm0}).
  Shown are the final equilibrium results for different thermodynamics treatments of the disk.
  The isothermal and adiabatic model have $H/r=0.037$ in accordance with the radiative model \citep[after][]{2009A&A...506..971K}.
 }
   \label{fig:kley-nelson-torque}
\end{center}
\end{figure}
The effect of this procedure on the resulting torque is shown in Fig.~\ref{fig:kley-nelson-torque} (left panel).
The basis for the four runs is the same equilibrium disk model constructed using all terms on the
right hand side of eq.~(\ref{eq:kley-nelson-energy}) and no planet. 
Embedding a planet of 20 $\Me$ yields {\it positive torques} only for radiative disks in the long run, 
where the maximum effect is given when only viscous heating and local radiative cooling are considered.
Obviously, the ability of the disk to exchange energy with its external environment
can maintain an entropy gradient. 
The strength of this positive corotation
effect also scales with the square of the planet mass up to about 20 -- 25 $\Me$, i.e. with the same
scaling as $\Gamma_0$.
Beyond a mass of $m_{\rm p} \approx 30 \; \Me$ gap opening begins, so that only Lindblad torques remain,
and the planets begin to slow their outward migration and eventually migrate inward.
In full 3D radiative simulations the results are qualitatively the same \citep{2009A&A...506..971K},
but interestingly they show an even stronger effect. Note that the full 3D models take into account 
all terms in eq.~(\ref{eq:kley-nelson-energy}) as the radiative diffusion acts along all spatial directions.

To analyse the spatial origin of the torques, it is useful to plot the radial torque
$d \Gamma(r)/dm $, which we define here  \citep[following][]{2010ApJ...724..730D},
such that the total torque $\Gamma_{\rm tot}$ is given as
\begin{equation}
\label{eq:gamm-dens}
            \Gamma_{\rm tot} = 2 \pi \int \frac{d \Gamma}{d m} (r) \, \Sigma(r) \, r dr\,.
\end{equation}
In other words, $d\Gamma(r)$ is the torque
exerted by a disk annulus of width $dr$ located at the radius $r$ and having the mass $dm$.
As $d \Gamma(r)/dm$ scales with the mass ratio squared and as $(H/r)^{-4}$, we rescale our results in units of
\begin{equation}
\label{eq:gamm0}
     \left( \frac{d \Gamma}{d m}\right)_0 =  \Omega_p^2(a_{\rm p}) a_{\rm p}^2  q^2 \left(\frac{H}{a_{\rm p}}\right)^{-4},
\end{equation}
where quantities with index $p$ are evaluated at the planet's position.
In Fig.~\ref{fig:kley-nelson-torque} (right panel) we show $d \Gamma(r)/dm $ for various disk models
all having $H/r=0.037$, as this is the value obtained from the radiative model.
The Lindblad contribution for the adiabatic and radiative simulations are nearly identical as to be expected
for optically thick disks. 
The `spike' at about $r \approx 0.99$ for the radiative model shows the positive contribution
due to the unsaturated corotation torque.

To capture the effect of the entropy-related horseshoe drag, approximate torque formulae have been
developed for adiabatic disks \citep{2009ApJ...703..857M,2010MNRAS.401.1950P}. 
However, these apply only for the initial unsaturated
torque, and care has to be taken when they are used in population synthesis models.
Similar to the vortensity-related torque, where viscosity prevents saturation, the entropy-related torque
will only be sustained in the presence of thermal diffusion. Optimal desaturation occurs
when $\tau_{\rm diff} \lesssim \tau_{\rm lib}$, where the diffusion time across the horseshoe region
is $\tau_{\rm diff}=x_{\rm s}^2/\kappa_{\rm T}$ and $\kappa_{\rm T}$ denotes the thermal diffusivity.
The effects of viscous and thermal diffusion have been considered simultaneously and refined expressions for
the torque have been constructed \citep{2010ApJ...723.1393M,2011MNRAS.410..293P}. 
A direct comparison of these torque formulae with full 3D radiative simulations shows clear discrepancies
\citep{2011A&A...536A..77B}.
This may be related to a physical difference between the full 3D simulations and the 2D results on which the
formulae are based. The inclusion of thermal diffusion in a 2D disk
operates only within the midplane, whereas for a real disk the cooling occurs along the vertical direction
and is balanced by internal dissipation or stellar heating.
Indeed, Fig.~\ref{fig:kley-nelson-torque} shows that torque reversal is strongest if only local cooling and no diffusion in
the midplane is considered. The radial range over which the total torque remains positive has been studied, and
\citet{2011A&A...536A..77B} find that corotation and Lindblad torques cancel each other at zero-torque
radii, which lie between 16 and 30 AU for planets with masses $20-30 \; \Me$. Such locations can act
as convergence points for growing planetary cores of similar mass.

\subsection{Massive planets: gap formation and type~II migration}
\label{subsect:type-II}
The analysis presented in the previous section refers to the situation of embedded low-mass planets that do not change the disk's original structure significantly. However, for larger planet masses the interaction becomes increasingly non-linear and the density profile in the disk will be modified.
In the following we describe the consequences of this process in more detail.
\subsubsection{Gap formation}
\label{subsubsec:gap-formation}
Upon increasing the planet mass, the strength of the gravitational interaction 
and the corresponding angular momentum transfer to the disk become stronger. 
If the angular momentum can be deposited locally in the disk and is not carried away by the spiral waves,
which may occur through viscous dissipation or shock waves,
the material inside (outside) the planet loses (gains) angular momentum and recedes from the planet.
Consequently, the material appears to be `pushed' away from the location of the planet and a gap begins to open in the disk.
The depth and width of the gap that the growing planet carves out will depend on the disk physics (viscosity and pressure), and on the
mass of the planet. The planet's transfer of angular momentum to the disk can be obtained 
by summing over the Lindblad resonances \citep{1980ApJ...241..425G} or by using the 
impulse approximation, where one considers the momentum change between a planet and individual disk 
particles shearing past \citep{1979MNRAS.186..799L}.
Both approaches yield similar results and \citet{1986ApJ...309..846L} give the 
following expression for the rate of angular momentum transfer
from the planet to the disk
\begin{equation}
\label{eq:kley-nelson-jdot-t}
   \dot{J}_{\rm tid} = f_{\rm tid} \, q^2 \Sigma_{\rm p} a_{\rm p}^2 \Op^2  \, \left(\frac{a_{\rm p}}{\Delta}\right)^3,
\end{equation}
with $f_{\rm tid} = 0.23$, and where $\Delta = |a_{\rm p} - r|$ denotes the impact parameter between the fluid
particle's unperturbed trajectory and the planet.
Viscosity opposes this and tries to close the gap; the corresponding viscous torque is
\citep{1974MNRAS.168..603L}
\begin{equation}
\label{eq:kley-nelson-jdot-v}
   \dot{J}_{\rm visc} = 3 \pi \Sigma_{\rm p} \nu a_{\rm p}^2 \Op,
\end{equation}
where $\nu$ is the vertically integrated kinematic viscosity.
Gap formation implies that the effect of gravity (the tidal torque) overwhelms that of viscosity,
i.e. $\dot{J}_{\rm tid} \gtrsim \dot{J}_{\rm visc}$. The gap should have a minimum
width at least equal to the size of the planet's Hill sphere (i.e. $\Delta \approx R_{\rm H}$),
this being $R_{\rm H} = (q/3)^{(1/3)} a_{\rm p}$.
The viscous criterion for gap formation is then given by \citep{1993prpl.conf..749L} 
\begin{equation}
\label{eq:kley-nelson-gap-v}
    q \gtrsim \frac{40 \nu}{a_{\rm p}^2 \Op}.
\end{equation}
Additionally, pressure effects in the disk tend to oppose gravitational influence of
the planet. For the disk response to be non-linear near the planet, so that the
spiral waves form shocks and deposit their angular momentum flux locally in the disk,
we require that the planet's Hill sphere size exceeds the disk thickness, leading to the {\it thermal}
gap opening criterion, $R_{\rm H} \gtrsim H$ \citep{1997Icar..126..261W}.

The combined action of viscosity and pressure have been analyzed in detail by \citet{2006Icar..181..587C}
and they quote the following criterion for the formation of a gap
\begin{equation}
\label{eq:kley-nelson-gap-crida}
    \frac{3}{4} \, \frac{H}{R_{\rm H}} + \frac{50 \nu}{q \, a_{\rm p}^2 \Op} \lesssim 1,
\end{equation}
which gives good agreement with full numerical simulations. 
The form and depth of the gap for different planet masses and viscosities is given, for example, in 
\cite{2002A&A...385..647D} and \cite{2006Icar..181..587C}. 
For solar nebula-type conditions, a Saturn mass planet begins to open visible gaps in the disk, but it should 
be kept in mind that gap formation is a continuous process.

An interesting side aspect is the analysis of nearly inviscid disks for which gap formation begins already for
small planetary masses. Under this circumstance the opening criterion depends of the disk mass as well 
\citep{2002ApJ...572..566R}, which has been confirmed by numerical simulations \citep{2009ApJ...690L..52L}.
Additionally, the process of gap formation has been analyzed analytically and numerically by a number
of researchers \citep{1994ApJ...421..651A,1996ApJ...460..832T,1999ApJ...514..344B,1999MNRAS.303..696K}.

\subsubsection{Type~II migration}
\label{subsubsec:type-II}
Because the density in the coorbital region is reduced, corotation torques are no longer
of any importance for larger planet masses. For very large masses when the planet has opened a gap
Lindblad torques are also reduced, resulting in a slowing down of the planet migration. In this case,
the planet is coupled to the viscous evolution of the disk \citep{1986ApJ...309..846L}. 
The migration timescale is given by the viscous diffusion time of the disk, $\tau_{\rm visc} \propto r_{\rm p}^2/\nu$. 
This non-linear regime of disk-planet interaction in which gap-opening planets are locked to the viscous evolution
is known as  {\it type~II} migration \citep[see also][]{1997Icar..126..261W}.

Using the above descriptions for the tidal and viscous torque (see eqs.~\ref{eq:kley-nelson-jdot-t} and \ref{eq:kley-nelson-jdot-v})
\citet{1986ApJ...309..846L} modeled the migration of a gap-opening massive planet simultaneously with the viscous disk evolution
utilizing the time-dependent diffusion equation for the disk surface density, $\Sigma(r,t)$, \citep{1981ARA&A..19..137P}.
Later, fully 2D hydrodynamical simulations of moving and accreting planets were performed by
\citet{2000MNRAS.318...18N}, who showed that a
Jupiter-mass planet starting from $\sim 5$~AU migrates on a timescale of about $10^5$ yrs all the way to the star.
If accretion can occur onto the planet, then its mass can grow up to 4-5 $\MJ$ during the migration.
Once the planet mass increases and the disk mass declines, i.e. when $\pi r_{\rm p}^2 \Sigma_{\rm p} \lesssim m_{\rm p}$,
the inertia of the planet begins to be important and `resists' the viscous driving, leading to a further reduction
in the speed of migration below the viscous diffusion \citep{1995MNRAS.277..758S,1999MNRAS.307...79I}, in agreement
with the numerical simulations \citep{2000MNRAS.318...18N}. 
The migration and accretion history for a range of planet masses has been analyzed by \citet{2008ApJ...685..560D}.

These results on migration have indicated a possible scenario to produce the observed population of hot Jupiters
\citep{1996Natur.380..606L,2002MNRAS.334..248A}. 
However, one has to keep in mind that the presence of additional massive planets in the system 
can lead to quite different behaviour with gravitational interaction and scattering processes between the planets
(see Sect.~\ref{sect:multi-planet}).
The presence of planets in disks opens up direct observational consequences. 
A gap could be detected directly by ALMA \citep{2005ApJ...619.1114W}, and possibly the accretion luminosity of  
the growing planet \citep{2006A&A...445..747K}. Lower mass planets create a gap in the dust distribution
\citep{2006A&A...453.1129P} or can cast shadows in the disk \citep{2009ApJ...700..820J}, both with observable consequences.

\subsection{Type~III (or runaway) migration}
\label{subsect:type-III}
Discussion so far has focused on the torque experienced by a
non migrating planet orbiting in a gaseous disk. A mode of
migration known as type~III or runaway migration, however, depends
on the radial migration of the planet and can generate periods of
very rapid migration under appropriate conditions
\citep{2003ApJ...588..494M, 2004ASPC..324...39A}. In principle,
type~III migration can be either inward or outward, although the
presence of Lindblad torques driving planets inward tends to bias
type~III migration to be inward in the absence of special conditions
such as very steep positive surface density gradients in the disk.

\subsubsection{Basic theory}

\begin{figure}[ht!]
\includegraphics[width=0.48\textwidth]{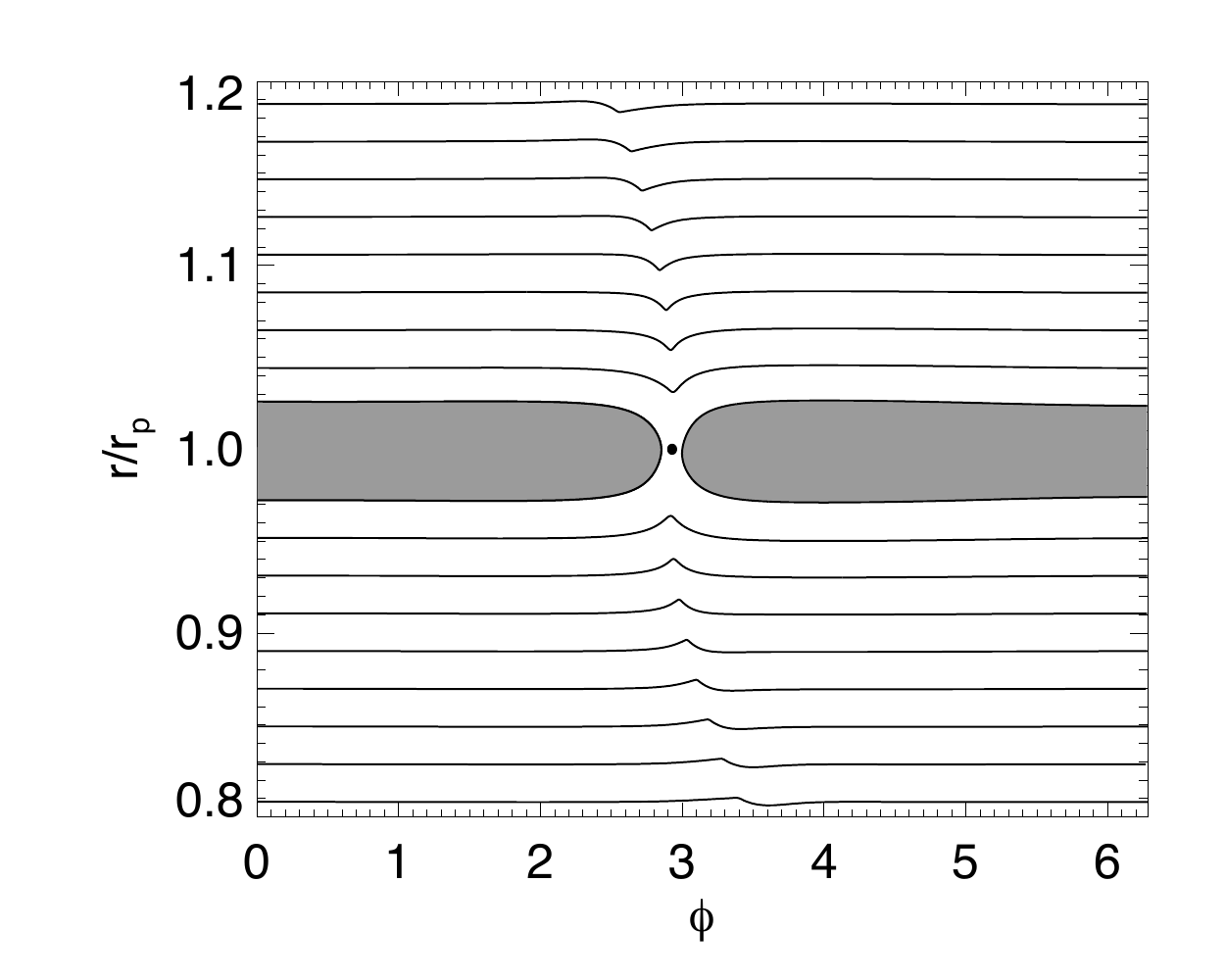}
\includegraphics[width=0.52\textwidth]{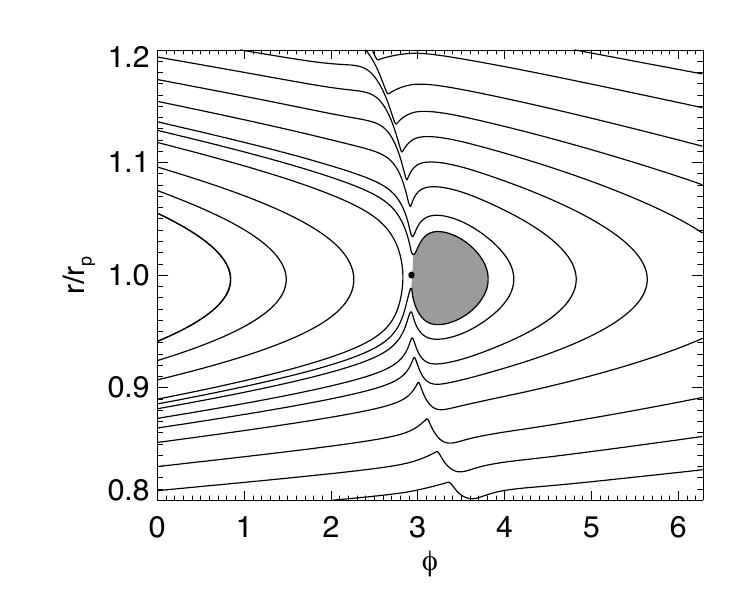}
 \caption{Streamlines for a $q=6 \cdot 10^{-5}$ (10 $\Me$) planet embedded in a disk with $H/r=0.05$
that illustrate how radial migration affects the flow topology (as viewed
in the planet reference frame).
The left panel corresponds to a non migrating planet
(compare to right panel of Fig.~\ref{fig:kley-nelson-topology}). 
The right panel shows a planet that is migrating inward on a timescale
of 300 orbits. A dramatic change in streamline topology is apparent
due to the rapid migration, as are the streamlines that connect directly between
the inner and outer disk. Shaded regions indicate librating material that
moves with the planet.}
\label{fig:streamlines-migrating}
\end{figure}

To simplify the discussion, we assume for the time being that the
previously discussed vorticity- and entropy-related corotation torques
are inactive. The Lindblad torque, $\Gamma^{\rm L}$, is assumed to operate
normally. Consider a planet undergoing inward radial migration on
a circular orbit. Material located in the vicinity of the separatrix between
inner disk material that is on circulating streamlines,
and material on librating horseshoe orbits, will undergo a single U-turn
encounter with the planet as it crosses the entire horseshoe region
and moves from the inner to the outer disk (see Fig.\ref{fig:streamlines-migrating},
where the right panel shows streamlines connecting the inner and outer disk for a rapidly migrating
planet).
As a result of this U-turn encounter, there is a change in the specific angular
momentum of the disk material equal to $\delta j = \Op a_{\rm p}^2  x_{\rm s}$.
This is just the change in specific angular momentum resulting from a fluid element moving
from an orbit located at radial distance $a_{\rm p} - x_{\rm s}$
to one located at $a_{\rm p} + x_{\rm s}$. The flux of mass passing through the inner
separatrix and encountering the planet is
${\dot M} = 2 \pi a_{\rm p} {\dot a}_{\rm p} \Sigma_{\rm s}$, where
${\dot a}_{\rm p}$ is the migration rate of the planet and
$\Sigma_{\rm s}$ is the surface density at the inner separatrix. The total
corotation torque exerted by the material flowing past the planet is therefore
\begin{equation}
\Gamma_{\rm Flow} = 2 \pi \Sigma_{\rm s} {\dot a_{\rm p}} \Op a_{\rm p}^3 x_{\rm s}.
\label{eq:torq_III}
\end{equation}
We refer to this as the `flow-through' corotation torque, and note that it depends on
the migration rate of the planet, such that there is a positive feedback between
the migration rate and the torque. In principle, this can lead to runaway migration
\citep{2003ApJ...588..494M}.

As the planet migrates inward, it carries with it material that is trapped in horseshoe
orbits and bound to the planet within its Hill sphere.
The planet must therefore exert a negative torque on this material (in which case the material
exerts a positive torque on the planet, acting as a drag by slowing its inward
migration). Denoting the mass trapped in the horseshoe region as $m_{\rm HS}$ and the
mass in the Hill sphere as $m_{\rm Hill}$, the drift rate of the planet plus trapped-fluid system,
migrating because of the flow-through and Lindblad torques, is given by
\begin{equation}
(m_{\rm p} + m_{\rm Hill} + m_{\rm HS}) \Op {\dot a}_{\rm p} a_{\rm p} = 
(4 \pi a_{\rm p}^2 x_{\rm s} \Sigma_{\rm s}) {\dot a_{\rm p}} \Op a_{\rm p} + 
2\Gamma^{\rm L}.
\label{eq:typeIII_drift}
\end{equation}
The quantity $(4 \pi a_{\rm p}^2 x_{\rm s} \Sigma_{\rm s})$ on the right-hand side
of eq.~(\ref{eq:typeIII_drift}) is the mass that would be contained in the horseshoe region
if the surface density there was equal to $\Sigma_{\rm s}$, the surface density at the inner
separatrix. The quantity $\delta m = 4 \pi a_{\rm p}^2 x_{\rm s} \Sigma_{\rm s} - m_{\rm HS}$
is therefore the approximate difference between the mass that would be contained in the
horseshoe region if the disk was unperturbed by the planet and the mass that is actually
contained in this region. $\delta m$ is often referred to as the 
{\it coorbital mass deficit} \citep{2003ApJ...588..494M}.
If the planet begins to form a gap in the disk, then $\delta m$ increases.

If we now treat the planet mass as being the sum of the actual planet
mass and the mass bound in the Hill sphere ($m_{\rm p}'=m_{\rm p}+m_{\rm Hill}$),
then eq.~(\ref{eq:typeIII_drift}) can be written as
\begin{equation}
{\dot a}_{\rm p} \Op a_{\rm p} (m_{\rm p}' -\delta m) = 2 \Gamma^{\rm L}\,,
\label{eq:typeIII_drift2}
\end{equation}
from which the drift rate is
\begin{equation}
{\dot a}_{\rm p} = \frac{2 \Gamma^{\rm L}}{\Omega_{\rm p} a_{\rm p} (m_{\rm p}' -\delta m)}.
\label{eq:typeIII_drift3}
\end{equation}
This equation demonstrates two important points. The first is that for
a planet that does not form a gap ($\delta m=0$), the migration rate is equal to that
due to Lindblad torques only. For an inwardly migrating planet, the negative flow-through
corotation torque is, therefore, balanced by the positive drag provided by the trapped
horseshoe material. The second point is that very rapid migration rates are clearly possible
when $\delta m \simeq m_{\rm p}'$, such that the coorbital mass deficit is similar to the
mass of the planet.
\begin{figure}[ht!]
\begin{minipage}{0.48\textwidth}
\includegraphics[width=0.95\textwidth]{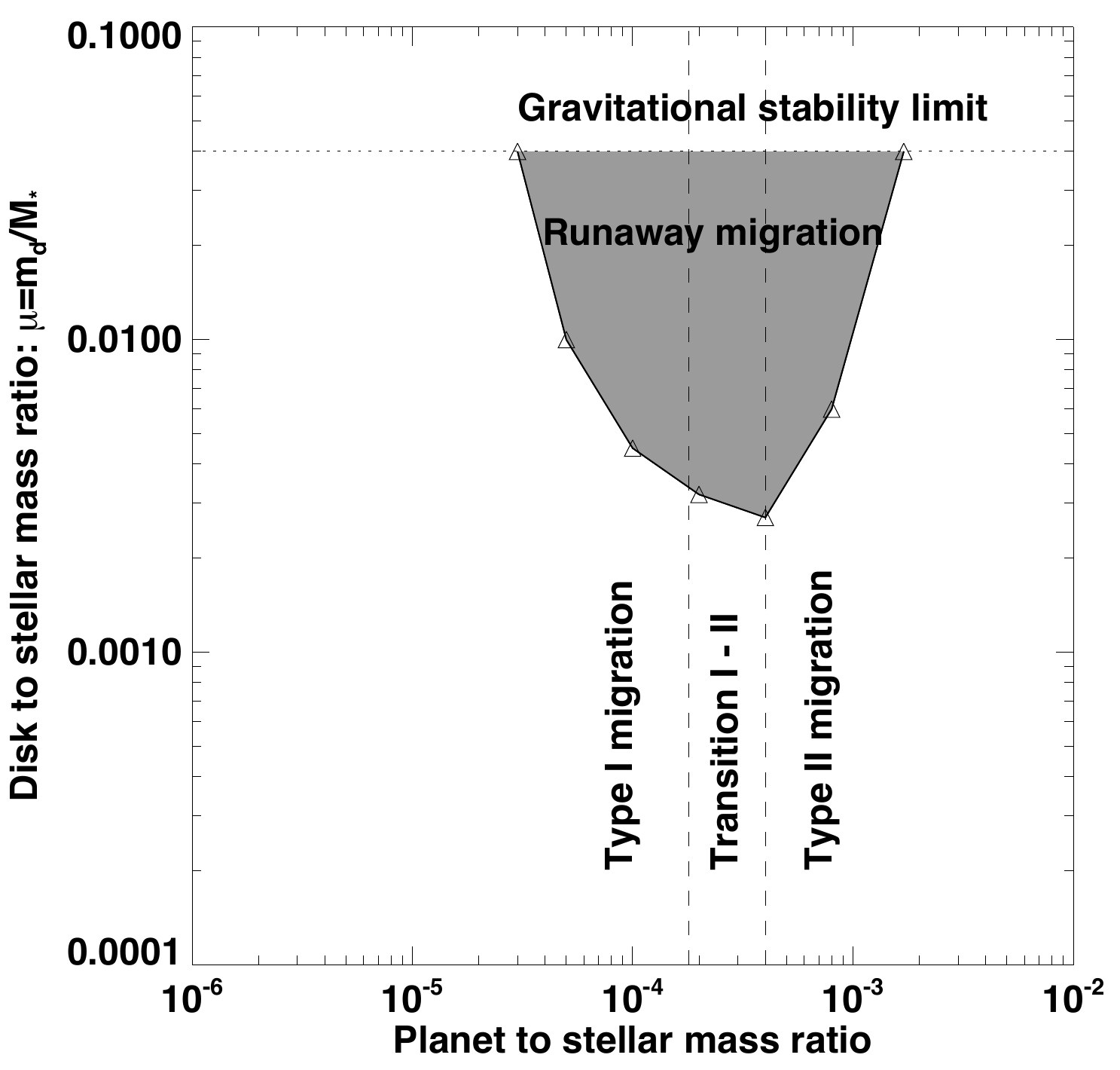}
\end{minipage}
\begin{minipage}{0.48\textwidth}
\includegraphics[width=0.95\textwidth]{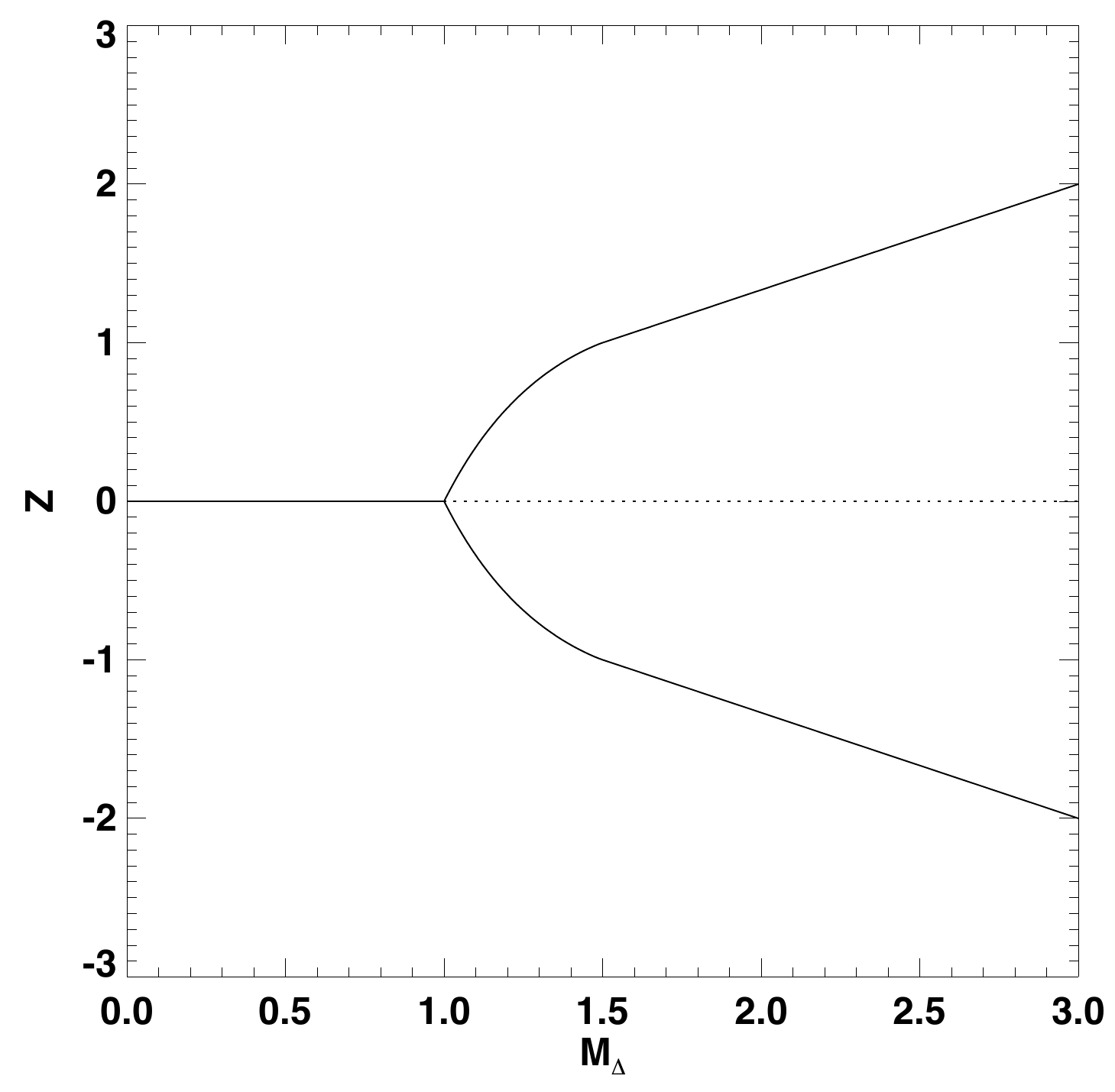}
\end{minipage}
\caption{
  The left panel shows planet mass versus the disk mass, and delimits the various
  types of migration behaviour in a disk with $H/r=0.04$ and dimensionless kinematic
  viscosity $\nu=10^{-5}$. Runaway or type~III migration can occur for
  parameters shown in the shaded region.
  The right panel illustrates the bifurcation behaviour predicted by eq.~(\ref{eq:typeIII_Z}) when $M_{\Delta}>1$,
  leading to rapid inward or outward type~III migration.}
\label{fig:TypeIII-conditions-bifurcate}
\end{figure}

According to eq.~(\ref{eq:typeIII_drift3}), the conditions required for rapid
migration are: ({\it i}) a partial gap must form at the planet location to generate a
coorbital mass deficit; ({\it ii}) the coorbital mass deficit $\delta m \simeq m_{\rm p}'$.
The mass of the planet expected to satisfy these conditions clearly depends on the
disk model, but for typical parameters rapid migration may occur for Saturn-mass planets
embedded in disks that are a few times more massive than the MMSN model.
Figure~\ref{fig:TypeIII-conditions-bifurcate} (left panel, adapted from \citet{2003ApJ...588..494M})
demonstrates the region of parameter space for which type~III migration occurs for
a disk with $H/R=0.04$ and dimensionless kinematic viscosity $\nu=10^{-5}$.

The above analysis is valid for migration rates that cause the planet to
migrate over a distance less than the horseshoe width, $a_{\rm p} x_{\rm s}$,
in one horseshoe libration time. We denote this migration rate as
${\dot a}_{\rm p,f}$. When this condition is violated then
eq.~(\ref{eq:typeIII_drift2}) should be modified 
\citep{2007prpl.conf..655P}.
When the Lindblad torque is small,  
solutions for steady inward or outward fast migration rates can be obtained. Setting $\Gamma^{\rm L}=0$,
and denoting $Z={\dot a}_{\rm p}/{\dot a}_{\rm p,f}$ and $M_{\Delta}=\delta m/m_{\rm p}'$,
one obtains
\begin{equation}
Z - \frac{2}{3} M_{\Delta} \, {\rm sign}(Z) \,  \left[1-(1-|Z|_{<1})^{3/2}\right] = 0,
\label{eq:typeIII_Z}
\end{equation}
where $|Z|_{\rm <1}={\rm MIN}(|Z|,\;1)$.
For $M_{\Delta} <1$, the only solution is $Z=0$. For $M_{\Delta} >1$, the solutions
bifurcate into rapid inward or outward migration as shown in
the right panel of Fig.~\ref{fig:TypeIII-conditions-bifurcate}.
An unstable solution corresponding to no migration ($Z=0$) also exists.
The fast migration solution, corresponding to migration rate ${\dot a}_{\rm p,f}$,
arises when $M_{\Delta}=3/2$.

\subsubsection{Simulation results}
A large suite of simulations examining the conditions for runaway migration
were presented by \citet{2003ApJ...588..494M}. These simulations produced
steady inward migration when $\delta m < m_{\rm p}'$
in disks with smooth surface density profiles $\Sigma(r) = \Sigma_0 r^{-3/2}$.
For models where $\delta m > m_{\rm p}'$, however, accelerating or runaway
migration was observed; and in extreme cases, rapid migration caused
the planet's semimajor axis to halve in less than 50 orbits. In all cases
examined, the runaway phase of migration eventually stalled.
Two factors prevent sustained runaway
migration over long times. The first is that the horseshoe region no longer forms a
closed system during rapid migration, so the coorbital mass deficit can
be lost. The second is that the factor $a_{\rm p}^2 \Sigma_{\rm s}$ that
enters the expression for the coorbital mass deficit is generally a decreasing
function of semimajor axis, such that $\delta m$ decreases as the planet moves inward,
eventually reaching a location where runaway migration can no longer occur
because $\delta m < m_{\rm p}'$. Allowing accretion onto the planet as it
migrates therefore causes runaway migration to stall earlier.

Numerical simulations presented by \citet{2008MNRAS.386..179P} display
rapid type~III migration for Jovian mass planets, and in basic agreement
with \citet{2003ApJ...588..494M}, migration that approximately halves the
initial semimajor axis in 50 orbits was obtained, followed by stalling
of the rapid migration. In a follow-on paper, \citet{2008MNRAS.387.1063P}
examine rapid outward migration and find an approximate doubling of planet
semimajor axes before the outward migration stalls after $\simeq 30$ orbits.
Outward migration in this case is initiated by setting-up a disk with a
low-density inner cavity with a sharp edge, and placing the planet close
to this edge such that it feels a strong and positive corotation torque
at the onset of the calculation.
\citet{2003ApJ...588..494M} also demonstrated the feasibility of rapid
outward migration, and similarly obtained an approximate factor of two
increase in the semimajor axis of the planet before the migration stalled
and reversed. Outward migration in this case was initiated by driving the
planet outward using an artificial positive torque and then allowing
the planet to evolve freely. As discussed above, spontaneous
outward migration is not observed in disks with smooth profiles, but instead
requires conditions that overcome the bias toward inward migration
generated by the Lindblad torques.

\subsubsection{Implications for planet formation}
As discussed above, runaway or type~III migration can occur
for Saturn-mass planets forming in disks that are a few times more massive
than the MMSN model. If we consider the formation of a gas giant planet
{\it via} the core accretion model, then one possible scenario is that
the core forms sufficiently early in the disk lifetime that the disk
is quite massive during the gas accretion stage. During the core formation
and early gas accretion stage, rapid type~I migration may be prevented
by the vorticity- and entropy-related torques discussed in Sect.~\ref{subsec:adi-rad}.
As the planet grows through steady gas accretion and it begins to form a gap,
the corotation torque preventing rapid type~I migration weakens significantly,
allowing inward migration to ensue. This transition occurs for $m_{\rm p} \gtrsim 30$
M$_{\oplus}$ for typical disk parameters. Continued mass growth up to a Saturn-mass ($\sim 100$ $\Me$)
may then allow a burst of runaway or type~III rapid inward migration to occur,
halving the semimajor axis in a few thousand years before the rapid
migration stalls.  The planet may then migrate inward on the type~II
timescale as it forms a deep gap and is locked to the viscous evolution
of the disk.

\subsection{Eccentricity and Inclination Evolution}
\label{subsect:orb-elem}
In addition to changing the semi-major axis, planet-disk interaction can modify the planetary
eccentricity ($e_{\rm p}$) and inclination ($i_{\rm p}$) as well. 
In this general case, the forces acting on the planet from the disk have to be evaluated 
with respect to the orbital plane of the planet, and the
changes in the orbital elements are then given in terms of normal, tangential, and radial forces \citep{1976AmJPh..44..944B}. 
As the radial and vertical position of the planet is no longer fixed, the disk forces vary over an orbital period 
such that time averages have to be taken to calculate the secular effect on the planet. 

\begin{figure}[ht!]
\begin{center}
\begin{minipage}{0.49\textwidth}
 \includegraphics[width=0.99\textwidth]{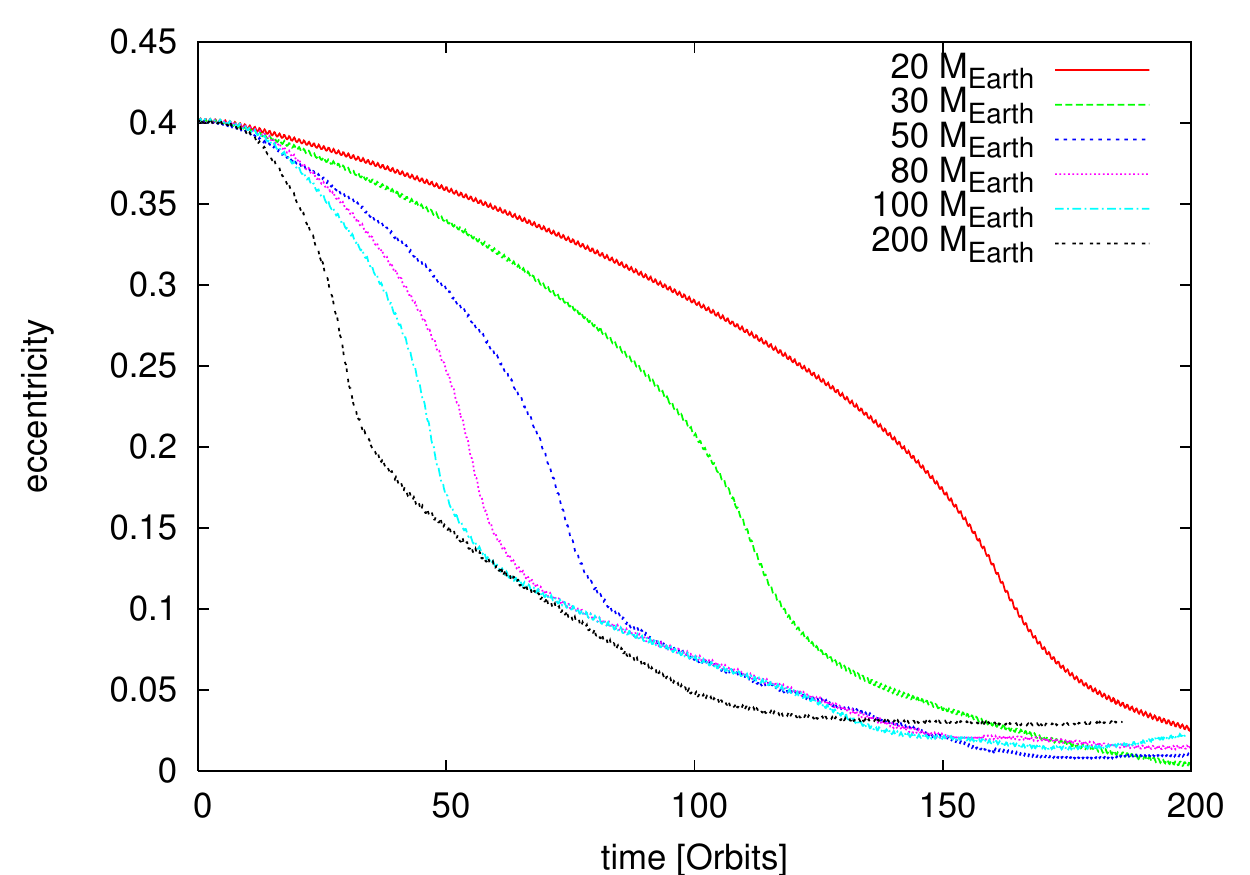}
\end{minipage}
\begin{minipage}{0.49\textwidth}
 \includegraphics[width=0.99\textwidth]{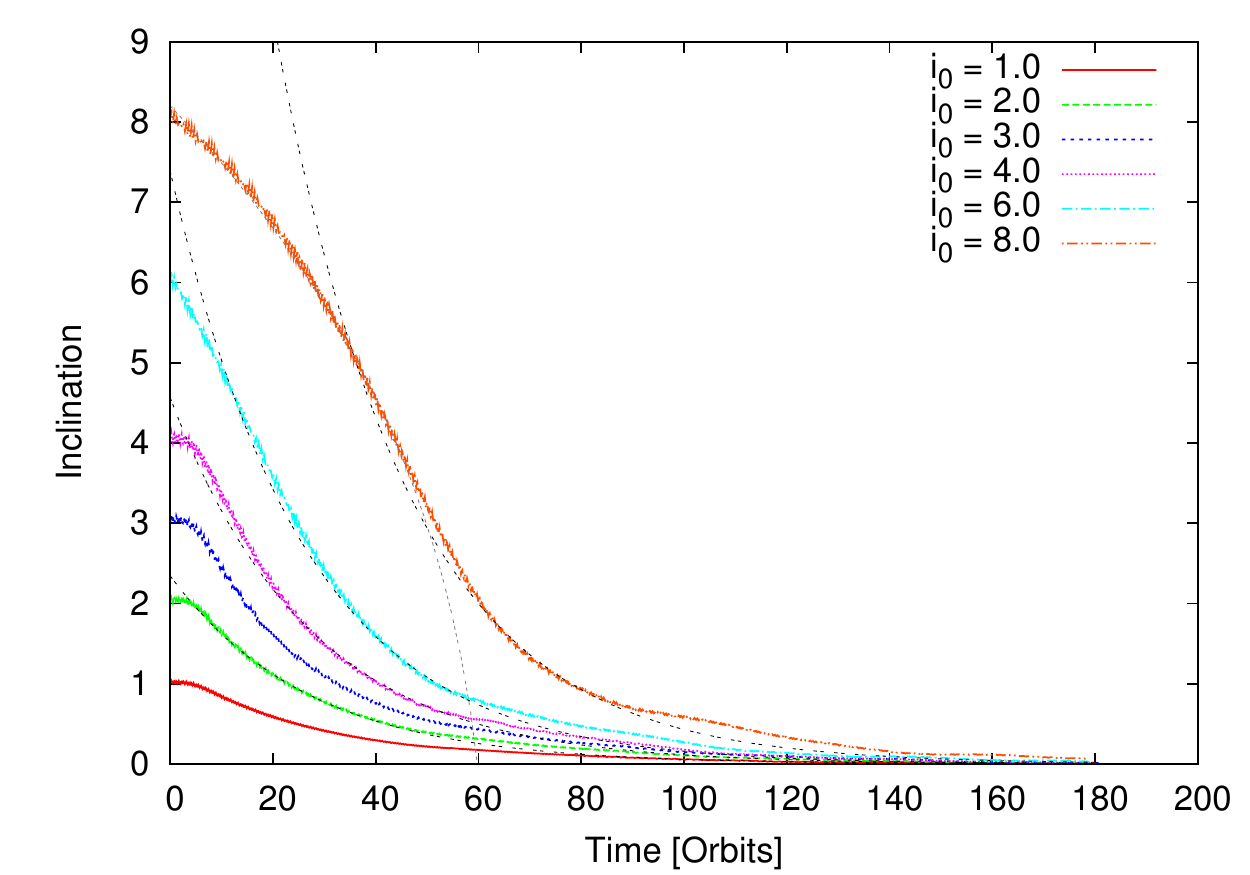}
\end{minipage}
 \caption{The change in orbital elements of planets embedded in 3D radiative disks. 
   {\bf Left:} Eccentricity change for planets with various masses starting with an initial $e_{\rm p}=0.4$.
   {\bf Right:} Inclination change for a $20 \; \Me$ planet starting with different initial inclinations.
    \citep[Plots adapted from][]{2010A&A...523A..30B,2011A&A...530A..41B}.
 }
   \label{fig:ecc-inc}
\end{center}
\end{figure}

\subsubsection{Eccentricity}
For smaller mass planets on eccentric orbits the linear analysis discussed in Sect.~2.1
can be adopted, but with an extended Fourier decomposition of the planet's 
gravitational potential \citep{1980ApJ...241..425G}
\begin{equation}
\psi_{\rm p}(r,\varphi,t) \, = \, 
      \sum_{m=0}^\infty \, \sum_{n=-m}^{+m} \psi_{m,n}(r) \cos\{m\varphi + (n-m) \Omega_{\rm p} t\}.
\label{eq:psi_ecc}
\end{equation}
Additional frequencies are now present in the problem compared to the circular orbit case because 
of epicyclic motion of the planet around its guiding centre. The pattern
speed associated with each potential component is $\Omega_{m,n}=(n-m)\Omega_{\rm p}/m$.
As with a planet on a circular orbit, Lindblad resonances occur at disk locations
where the perturbation frequency experienced by a fluid element equals the local
epicylic frequency, and corotation resonances occur wherever the pattern
speed equals the local angular velocity of the disk. For each value of $m$ and $n$ there
are three resonance locations \citep{1980ApJ...241..425G,2003ApJ...585.1024G}:
external Lindblad resonances that act to increase
the planetary eccentricity, corotation resonances that act to damp it (if unsaturated),
and coorbital Lindblad resonances that damp $e_{\rm p}$ 
\citep[see also][]{1986Icar...67..164W,1993ApJ...419..166A}.
Linear analysis for low-mass planets indicate rapid exponential damping of eccentricity,
$d e_{\rm p}/dt \propto - e_{\rm p}$, on timescales of $\tau_{\rm ecc} \approx (H/r)^2 \tau_{\rm mig}$ for small initial eccentricities
\citep{2004ApJ...602..388T}, primarily because of the damping provided by the
coorbital Lindblad resonances.
For larger initial eccentricity, with $e_{\rm p} \gtrsim H/r$, \citet{2000MNRAS.315..823P} find a different damping
behaviour with $d e_{\rm p}/dt \propto - e_{\rm p}^{-2}$. These results have been confirmed through 2D and  3D hydrodynamical simulations
of embedded low-mass planets up to 30 $\Me$ \citep{2007A&A...473..329C}, see also Fig.~\ref{fig:ecc-inc}, left panel.

The discussion concerning the influence of resonances for eccentric planets demonstrates
that the evolution of $e_{\rm p}$ depends on a balance between positive and negative
contributions. A fully embedded planet has its eccentricity damped largely
because of coorbital Lindblad resonances \citep{1993ApJ...419..166A}, but if the
planet opens a gap, then these are rendered inoperative. In this case, the question
of whether or not $e_{\rm p}$ increases is determined by the balance between
external Lindblad resonances and corotation resonances (that may be fully or partially saturated).
For fully unsaturated corotation resonances the eccentricity is expected to damp,
but calculations by \cite{2003ApJ...585.1024G} and \cite{2008Icar..193..475M}
suggest disk-driven eccentricity growth may be possible through (partial) saturation
of the corotation resonances \citep{2003ApJ...587..398O}.
This scenario, which should apply for planet masses around Saturn's mass, 
seems appealing to explain the high mean eccentricity of extrasolar planets. 
However, recent multi-dimensional studies of embedded planets of larger masses up to 1 $\MJ$ \citep{2010A&A...523A..30B} suggest that eccentricity
is damped quite generally, see Fig.~\ref{fig:ecc-inc} (left panel).
This damping behaviour is in very good agreement with
the low eccentricities found in some observed resonant planetary systems \citep{2002ApJ...567..596L}.
Indeed, the action of an (one-sided) inner disk on an outer planet results in the damping of
planetary eccentricity \citep{2008A&A...483..325C}. 

For massive planets of a few $\MJ$ a new but related
effect occurs in that the outer disk can become eccentric even
if the planet is on a circular orbit \citep{2001A&A...366..263P, 2006A&A...447..369K}. 
This phenomenon is also controlled by a balance between competing resonances. The instability 
has been observed to occur for
wide and deep gaps such that the eccentricity damping corotation resonance that coincides with the
outer 2:1 Lindblad resonance is cleared and the eccentricity driving 3:1 resonance
is dominant. The induced disk eccentricity can react back onto the planet and excite a non-zero $e_{\rm p}$. 
This  effect operates strongly for massive planets above 10 $\MJ$ \citep{2001A&A...366..263P}.
More recently, it has been reported to operate for Jovian-mass planets \citep{2006ApJ...652.1698D},
resulting in more modest $e_{\rm p}$ growth in this case.
Interestingly, for an eccentric disk the planet moves periodically into the disk, which allows for continued mass
accretion onto the planet \citep{2006A&A...447..369K}. This may be a mechanism to explain some of the large planet
masses ($\gtrsim 6 \MJ$) that are otherwise difficult to reach due to the onset of gap formation.

\subsubsection{Inclination}
The change in the planet's inclination has been studied in linear theory for small planet masses 
and small inclinations $i_{\rm p} \lesssim H/r$.
Exponential damping has been found, $d i_{\rm p}/dt \propto - i_{\rm p}$, which occurs on timescales $(H/r)^{2}$ 
times shorter than the migration timescale,
quite similar to the eccentricity damping time \citep{2003AJ....125.3389W,2004ApJ...602..388T}. 
This behaviour has been verified in 3D isothermal simulations by \citet{2007A&A...473..329C}. 
For larger inclinations $i_{\rm p} \gtrsim H/r$,
these researchers find inclination damping on a longer timescale with a behaviour $d i_{\rm p}/dt \propto - i_{\rm p}^{-2}$, which is 
the same scaling obtained for eccentricity damping when $e_{\rm p}$ is large.
These results have been verified in 3D radiative simulations by \citet{2011A&A...530A..41B}.
Additionally, the inclination has been shown to damp by disk-planet interaction 
for all planet masses up to about 1 $\MJ$ \citep{2009ApJ...705.1575M,2011A&A...530A..41B}, see Fig.~\ref{fig:ecc-inc} (right panel).

Due to the much shorter timescales of eccentricity and inclination damping in contrast to the migration time, it is to be expected that
for isolated planets embedded in disks these quantities should be very small, at least in laminar disks, i.e. even
if planets formed in disks with $e_{\rm p} \neq 0$ and $i_{\rm p} \neq 0$, they would be driven rapidly
to $e_{\rm p}=0$ and $i_{\rm p} =0$.

\section{Planets in turbulent disks}
\label{sect:turbulence}
Disk viscosity plays an important role in controlling the
dynamics of embedded planets: saturation of
corotation torques for low-mass planets and the structure
of the gap that forms in the presence of a giant planet are two
examples where viscosity provides a controlling influence.
The discussion presented in previous sections focused on laminar
disks, where anomalous viscous stresses are
provided through the Navier-Stokes equation combined with a
prescribed viscosity.
Viscous stresses in protoplanetary disks, however, most likely
arise because of MHD turbulence generated by the
magnetorotational instability (MRI) \citep{1991ApJ...376..214B}.
In this section, we discuss current understanding of how
disk turbulence influences the dynamics of embedded planets.

Understanding the role of disk turbulence in the context of
planetary formation and dynamics is hindered at present by an incomplete
understanding of MHD turbulence in protoplanetary disks. 
In the ideal MHD limit of a fully ionised disk, the strength of 
turbulence and the amplitudes of density and velocity fluctuations are strong 
functions of the net magnetic field strength and topology. Externally 
imposed vertical fields  generate more vigorous turbulence than toroidal 
fields, and disks that only host fields generated internally
through dynamo action yield the weakest turbulent flows 
\citep{1995ApJ...440..742H,2002ApJ...571..413S}. 
Neither the strengths nor geometries of magnetic fields in accretion disks
are well constrained by observations. The
high densities and low temperatures typical of protoplanetary disks
mean that in planet-forming regions (approximately
$0.5 \le R \le 20$ AU from the central star), material near the
midplane is likely to be insufficiently ionised to sustain MHD turbulence
(fractional gas-phase electron abundances $x_{\rm e} \lesssim 10^{-11}$).
Instead, protoplanetary disks are expected to have
a layered structure in which the surface layers are ionised
by external sources such as cosmic rays or X-rays, and the
shielded midplane region maintains a `dead zone' that remains in a
near-laminar state because of large Ohmic resistivity there
\citep{1996ApJ...457..355G}. Details of this layered structure depend on a
large number of parameters including the sizes and abundances of dust grains
\citep{2000ApJ...543..486S, 2006A&A...445..205I}, and the intensity of
external ionisation sources \citep{2009ApJ...703.2152T}, in addition to the
field strength and topology. Disk structure can also be influenced
by non-ideal MHD processes such as the Hall effect  and ambipolar diffusion
\citep{1999MNRAS.307..849W,2001ApJ...552..235B}, increasing substantially the
complexity and computational
expense of simulating the non linear evolution of such disks. It is for
this reason that until recently most studies of planets embedded in turbulent
disks have focused on disks without dead zones.

\subsection{Low-mass embedded planets}
Early studies of the interaction between low-mass, non gap-forming
planets and turbulent disks are presented by
\citet{2004MNRAS.350..849N}, \citet{2004MNRAS.350..829P} and
\citet{2004ApJ...608..489L}. Using 3D ideal MHD simulations
of turbulent disks with embedded low-mass planets,
\citet{2004MNRAS.350..849N} showed that disk torques
experienced by the planets included a stochastic component,
whose r.m.s. value was significantly larger in magnitude than
type I torques for planets with masses $m_{\rm p} \le 10 \; \Me$,
suggesting that these planets would experience a random walk
element in their migration. \cite{2004ApJ...608..489L} reached similar conclusions using
hydrodynamic simulations with forced turbulence excited
through a stirring potential fitted to the results of MHD simulations.
The orbital evolution of planets and planetesimals
in turbulent disks was considered by \cite{2005A&A...443.1067N}, who
demonstrated that embedded low-mass planets do indeed experience
a random walk in semimajor axis, and also experience growth in
their eccentricities due to interaction with the turbulent density
fluctuations.

These studies were unable to determine the long term evolution of planets due
to the computational expense of running large MHD simulations, and could run
for only $\lesssim 100$ planet orbits. As such, they were unable to address the
question of whether or not stochastic torques could dominate over type~I
torques during long evolution times and prevent the rapid inward migration of low-mass
planets. Considering a simple discrete random walk model in which planets
receive positive and negative kicks, $\delta J$, in their angular momenta,
the cumulative change in angular momentum after $N$ kicks is expected to
be $\Delta J = \delta J \sqrt{N}$. Denoting the time that elapses between
kicks as $t_{\rm corr}$ (the correlation time associated with the stochastic
torques), the total time elapsed is $t=N \; t_{\rm corr}$.
Denoting the r.m.s. amplitude of the stochastic torques by $\sigma_{\rm T}$ gives
$\delta J= \sigma_{\rm T} t_{\rm corr}$ and 
\begin{equation}
\Delta J = \sigma_{\rm T} t_{\rm corr} \sqrt{\frac{t}{t_{\rm corr}}}.
\label{eq:Delta_J}
\end{equation}
The change in angular momentum induced by type~I torques of magnitude $\Gamma_{\rm I}$
in time $t$ is $\Delta J = \Gamma_{\rm I} \; t$.  Therefore, we expect type~I
and stochastic torques to induce similar angular momentum changes after a time
given by \citep{2004MNRAS.350..849N},
\begin{equation}
t \sim \left(\frac{\sigma_{\rm T}}{\Gamma_{\rm I}}\right)^2 t_{\rm corr}.
\label{eq:t_converge}
\end{equation}
For a 1 $\Me$ planet embedded in a fully turbulent disk typical values are
$\sigma_{\rm T}/\Gamma_{\rm I} \simeq 50$ and $t_{\rm corr} \simeq 0.5$ planet orbits
\citep{2005A&A...443.1067N},
such that run times of $> 1,000$ orbits are required to observe torque convergence.

\begin{figure}[ht!]
\begin{center}
\begin{minipage}{0.49\textwidth}
\includegraphics[width=0.99\textwidth]{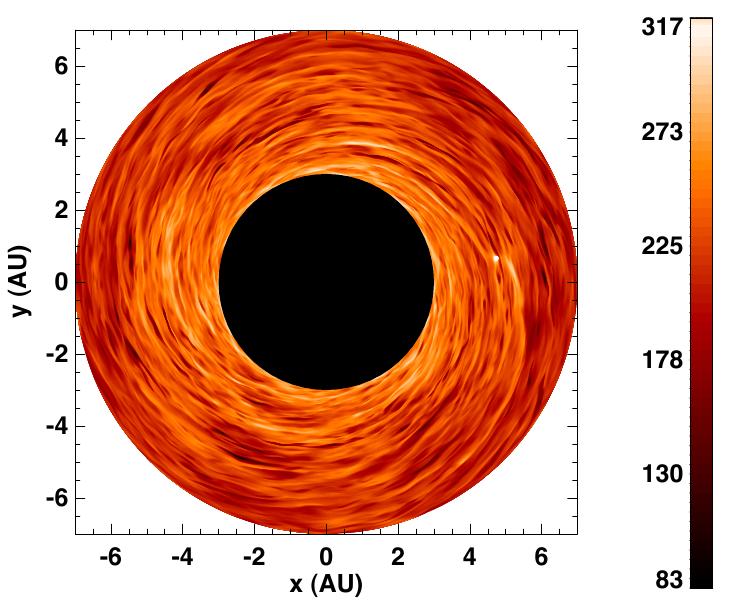}
\end{minipage}
\begin{minipage}{0.50\textwidth}
\includegraphics[width=0.99\textwidth]{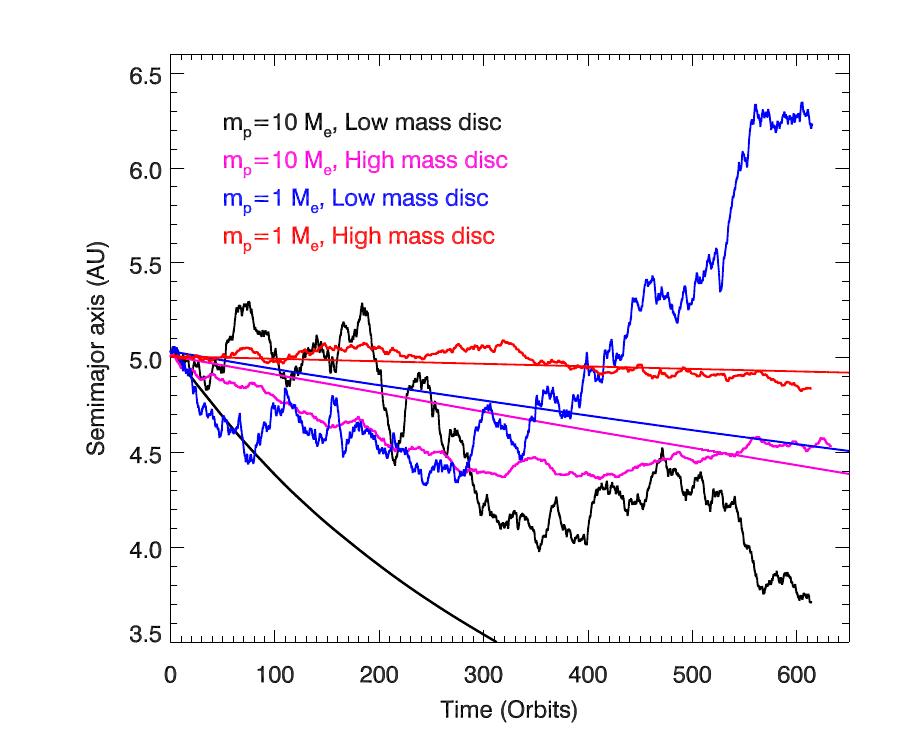}
\end{minipage}
 \caption{{\bf Left}: Snapshot of surface density (g cm$^{-2}$) from a simulation with a $10 \, \Me$ planet 
  embedded in a 3D  turbulent disk  with  H/R=0.05  and effective  $\alpha \simeq 0.015$.
  {\bf Right}: The evolution of the planet semimajor axes
  for turbulent models with different planet and disk masses. The smooth lines show
  the evolution of planets in equivalent laminar disks.}
\label{fig:turb-figs}
\end{center}
\end{figure}

To illustrate the results of MHD simulations with
embedded planets we present the results of four simulations in
Fig.~\ref{fig:turb-figs}, where the disk set-up is as described in
\citet{2010MNRAS.409..639N} with $H/R=0.05$. Stresses generated
by the turbulence give $\alpha \simeq 0.015$ throughout the disk. The four simulations
have planet masses $m_{\rm p} \in \{1, \; 10\}$ $\Me$, and surface density at the initial
planet location ($r_{\rm p}= 5$ AU) of $\Sigma_{\rm p} \in \{180, \; 900\}$ g~cm$^{-2}$.
A snapshot from the run with a $m_{\rm p}=10 \; \Me$ planet
embedded in the lower mass disk is shown in the left panel of Fig.~\ref{fig:turb-figs}.
The planet wakes are just visible against the background turbulent fluctuations,
providing a vivid illustration of why turbulence is able to affect the orbital
dynamics of embedded planets. 
The right panel of
Fig.~\ref{fig:turb-figs} displays the semimajor axis evolution of the four planets,
in addition to migration trajectories of equivalent laminar disk simulations.
Orbital eccentricities are also excited by the turbulence, and for the
$m_{\rm p}=1 \; \Me$ cases, values of $e_{\rm p} \simeq 0.02$ are obtained
in the simulations. 
Each model shows clear influence of the turbulence affecting orbital evolution.

The simulations described above provide a degree of insight into
the role of turbulence in driving the long term evolution of planet orbits,
but this is ultimately a statistical problem in which the evolution of an
ensemble of planets should be considered.
An individual simulation provides a unique evolution history for a single
planet, but offers limited information about the evolution of a
planet population. Ideally, a large number of independent simulations should be
used to build-up a statistical picture of the distribution of outcomes, but
the computational expense is prohibitive at present.

An alternative approach that captures exactly this statistical nature of the problem,
and allows the long-term evolution to be examined,
has been presented by \citet{2006ApJ...647.1413J} and \citet{2009ApJ...701.1381A}.
Their studies examined the problem of type~I migration in the presence
of stochastic torques using a Fokker-Planck formalism. This results in an
advection-diffusion equation that governs the probability, $P(J,t)$, of
there being a planet with angular momentum $J$ at time $t$:
\begin{equation}
\frac{\partial }{\partial t} P(J,t) = \frac{\partial }{\partial J} 
\left[ \Gamma_{\rm I} P(J,t)\right] +
\frac{\partial^2 }{\partial J^2} \left[D_{\rm J} P(J,t) \right],
\label{eq:advection-diffusion}
\end{equation}
where $\Gamma_{\rm I}$ is the underlying type~I torque and
$D_{\rm J}$ is the diffusion coefficient that determines the
rate at which planet angular momentum random walks due
to stochastic torques. Given an initial distribution of planet
orbits, eq.~(\ref{eq:advection-diffusion})
describes how the distribution evolves over time.
The diffusion coefficient, $D_{\rm J}$, can be approximated
by $D_{\rm J} \simeq \sigma^2_{\rm T} t_{\rm corr}$, this being
consistent with eq.~(\ref{eq:Delta_J}). Given values for the r.m.s. 
fluctuating torque amplitude and correlation time of the stochastic component,
the problem is well defined (subject to the assumption that the torques can be
described as a simple superposition of type~I and stochastic components).

\cite{2006ApJ...647.1413J} expressed the torque r.m.s. in terms of
the gravitational force experienced by a particle suspended just above a flat 2D
disk with underlying surface density ${\overline \Sigma}$
(namely $2 \pi G {\overline \Sigma)}$:
\begin{equation}
\sigma_{\rm T} = C_{\rm D} 2 \pi G \, {\overline \Sigma} \, a_{\rm p}\,,
\label{equation}
\end{equation}
where $C_{\rm D}$ is an unknown constant that can be determined from simulations.
The value of $C_{\rm D}$ obtained from the simulations shown in Fig.~\ref{fig:turb-figs}
is $C_{\rm D}=0.035$, similar to the fiducial values of 0.046 adopted by
\citet{2006ApJ...647.1413J} and 0.05 adopted by \citet{2009ApJ...701.1381A}.
The correlation time derived from the simulations is $t_{\rm corr}=0.45$
[see the discussion in \citet{2010MNRAS.409..639N} concerning calculation
of correlation times, and the range of correlation times obtained in simulations], 
again similar to the value $t_{\rm corr}=0.5$ adopted in the
above studies.

\begin{figure}[ht!]
\begin{center}
\includegraphics[width=0.6\textwidth]{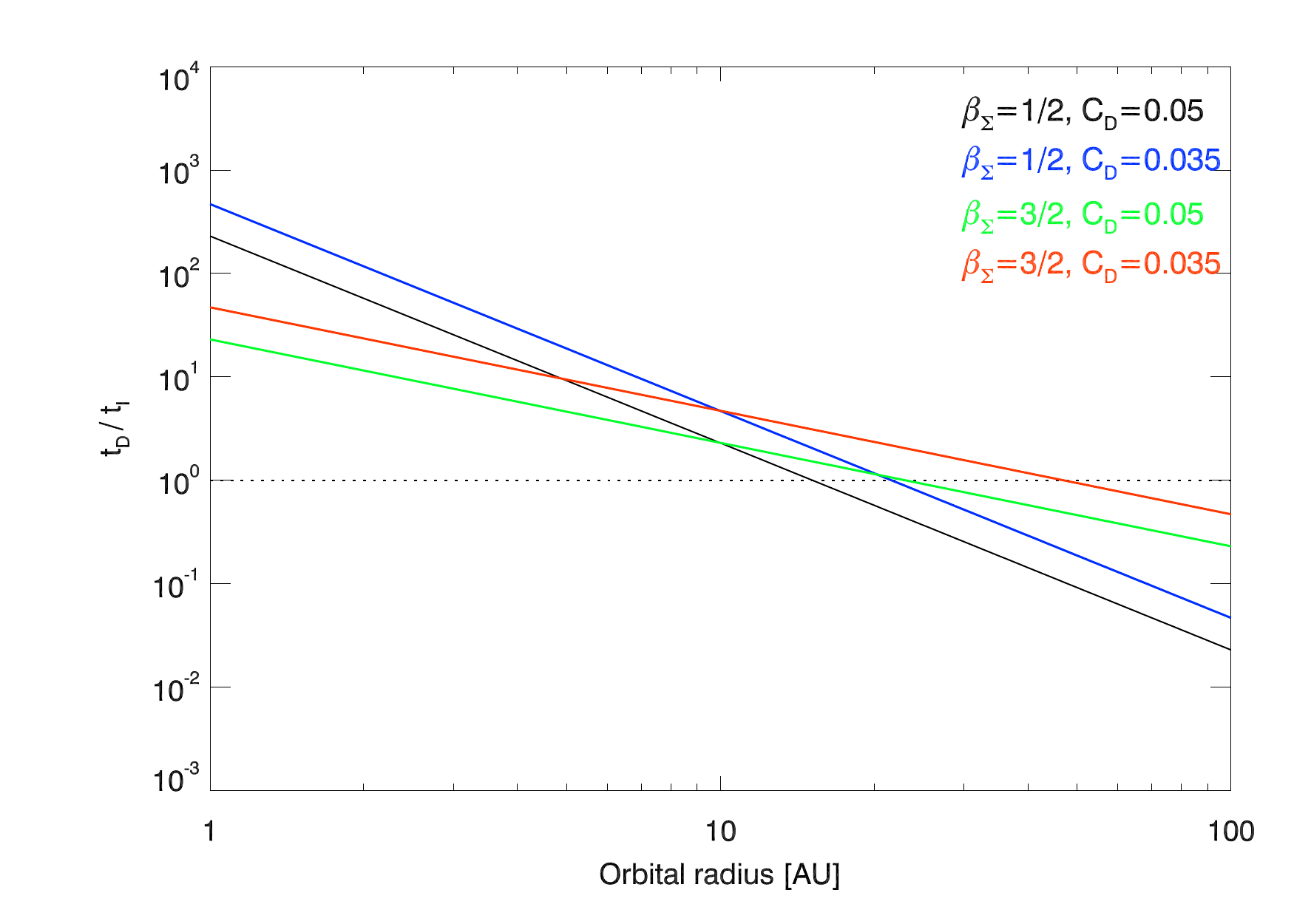}
 \caption{
This figure shows the ratio $t_{\rm I}/t_{\rm D}$ for
the two different disk models described in the text, and for two different values of
the turbulence parameter $C_{\rm D}$. The various curves fall beneath the value of
unity in the outer disk, indicating that diffusive migration dominates type~I
migration there.}
\label{fig:Tmig_Tdiff}
\end{center}
\end{figure}

The long-term evolution of planetary orbits is determined by the relative
values of the diffusion time $t_{\rm D} = J^2/D_{\rm J}$ and the
type~I migration time, $t_{\rm I} = J/\Gamma_{\rm I}$. For a fully
turbulent disk with a radially uniform value of the effective viscosity
parameter, $\alpha$, it is reasonable to suppose that the mean relative
surface density fluctuations, $\delta \Sigma/ {\overline \Sigma}$, are
constant, such that the parameter $C_{\rm D}$ defined above is
constant also. The ratio of diffusion to type~I timescale is
\begin{equation}
\frac{t_{\rm D}}{t_{\rm I}} = \frac{J  \, \Gamma_{\rm I}}{C_{\rm D}^2 
                              (2 \pi G {\overline \Sigma} a_{\rm p})^2},
\label{eq:tD_over_tI}
\end{equation}
where evolution dominated by stochastic torques requires $t_{\rm D} < t_{\rm I}$.
Considering eq.~(\ref{eq:tD_over_tI}) we note that type~I torques depend linearly
on ${\overline \Sigma}$, so that the influence of
stochastic torques is increased relative to type~I torques for a higher surface
density, ${\overline \Sigma}$. This is a prediction that has yet to be confirmed
satisfactorily by numerical simulations because of the need to run a large
ensemble of models.

Assuming that the disk surface density and temperature obey radial power laws with
indices equal to $\beta_\Sigma$ and $\beta_T$, respectively,
the ratio of timescales given by
eq.~(\ref{eq:tD_over_tI}) obeys the following proportionality relation
\begin{equation}
\frac{t_{\rm D}}{t_{\rm I}} \propto m_{\rm p} C_{\rm D}^{-2} 
a_{\rm p}^{(\beta_\Sigma + \beta_{\rm T} - 3)}.
\label{eq:tD_tI-proportion}
\end{equation}
As expected, increasing the planet mass increases the influence of
type~I migration. Increasing the value of $C_{\rm D}$ increases
the influence of stochastic torques and reduces $t_{\rm D}/t_{\rm I}$.
In most disk models, the power-law indices have values $0 \le \beta_\Sigma \le 3/2$ and
$1/2 \le \beta_{\rm T} \le 1$, such that the influence of diffusion increases as one moves 
to larger radii in a disk. 
This point is illustrated in Fig.~\ref{fig:Tmig_Tdiff}, in which we
plot the ratio $t_{\rm D}/t_{\rm I}$
for two disk models. The first model is five times more massive than
the MMSN with
$\Sigma(R) =\Sigma_0 (R/1 {\rm AU})^{-3/2}$ and $\Sigma_0=8,500$ g~cm$^{-2}$.
The second model has $\Sigma(R)=\Sigma_0 (R/1 {\rm AU})^{-1/2}$ with
$\Sigma_0=641$ g~cm$^{-2}$, and both models adopt $T(R)=T_0 (R/1 {\rm AU})^{-1/2}$
with $T_0=270$ K. Clearly, the region where stochastic torques
dominate over type~I torques ($t_{\rm D} / t_{\rm I} < 1$)
is in the outer disk beyond 10 AU for a disk that sustains
MHD turbulence throughout its interior with effective $\alpha \simeq 0.01$
[see also \citet{2006ApJ...647.1413J} and \citet{2009ApJ...701.1381A}].

\subsubsection{Effect of a dead zone}
To date, there have been no global dynamical studies of planets embedded in disks with
dead zones that are capable of measuring net migration torques. Local shearing box
simulations of protoplanets and planetesimals embedded in disks
with dead zones have been presented however \citep{2007ApJ...670..805O,2011MNRAS.415.3291G}.
These indicate that embedded bodies experience significantly reduced stochastic torques.
In a recent study, the dynamics of planetesimals embedded in disks with dead zones of different
depths were examined by \cite{2011MNRAS.415.3291G}.
For disk models similar to those used to generate Fig.~\ref{fig:Tmig_Tdiff}, the reduction
in stochastic torque amplitude, $\sigma_{\rm T}$, is at least one order of magnitude
when comparing a fully active disk with an equivalent one containing a dead zone.
The ratio of diffusion to migration times $t_{\rm D}/t_{\rm I} \propto \sigma_{\rm T}^{-2}$
(the diffusion coefficient determining the rate of stochastic migration scales
with the square of the r.m.s. stochastic torque),
such that even the most optimistic model plotted in Fig.~\ref{fig:Tmig_Tdiff}
would predict that type~I torques will dominate at all disk radii interior to 100 AU.
It is interesting to note, however, that recent models of protoplanetary
disks indicate that the dead zone may extend up to a radius of $\sim 15$ AU
\citep{2009ApJ...703.2152T} or only to $\sim 5$ AU \citep{2012MNRAS.420.2419F},
such that turbulence can still play an important role in the outer disk.

\subsection{Corotation torques}
\label{subsubsec:turb-corot}
Viscous diffusion is required to prevent saturation of both the vorticity- and
entropy-related contributions to the corotation torque (see Sect.~\ref{subsubsec:corot-sat}).
A key question is whether or not MHD turbulence acts in a similar way to the
anomalous viscosity used in most studies of the corotation torque, and 
prevents saturation. This issue was examined by \cite{2010ApJ...709..759B} 
using 2D hydrodynamic simulations in which turbulence was forced using the 
prescription proposed by \cite{2004ApJ...608..489L}.
Their study suggested that Lindblad and corotation torques
in turbulent disks converge to values similar to those obtained
in viscous laminar disks for planets with masses equal to
a few Earth masses, at least for disks in which the viscous $\alpha$
value was relatively small ($\alpha \lesssim 10^{-3}$).

\begin{figure}[ht!]
\begin{center}
\begin{minipage}{0.49\textwidth}
\includegraphics[width=0.99\textwidth]{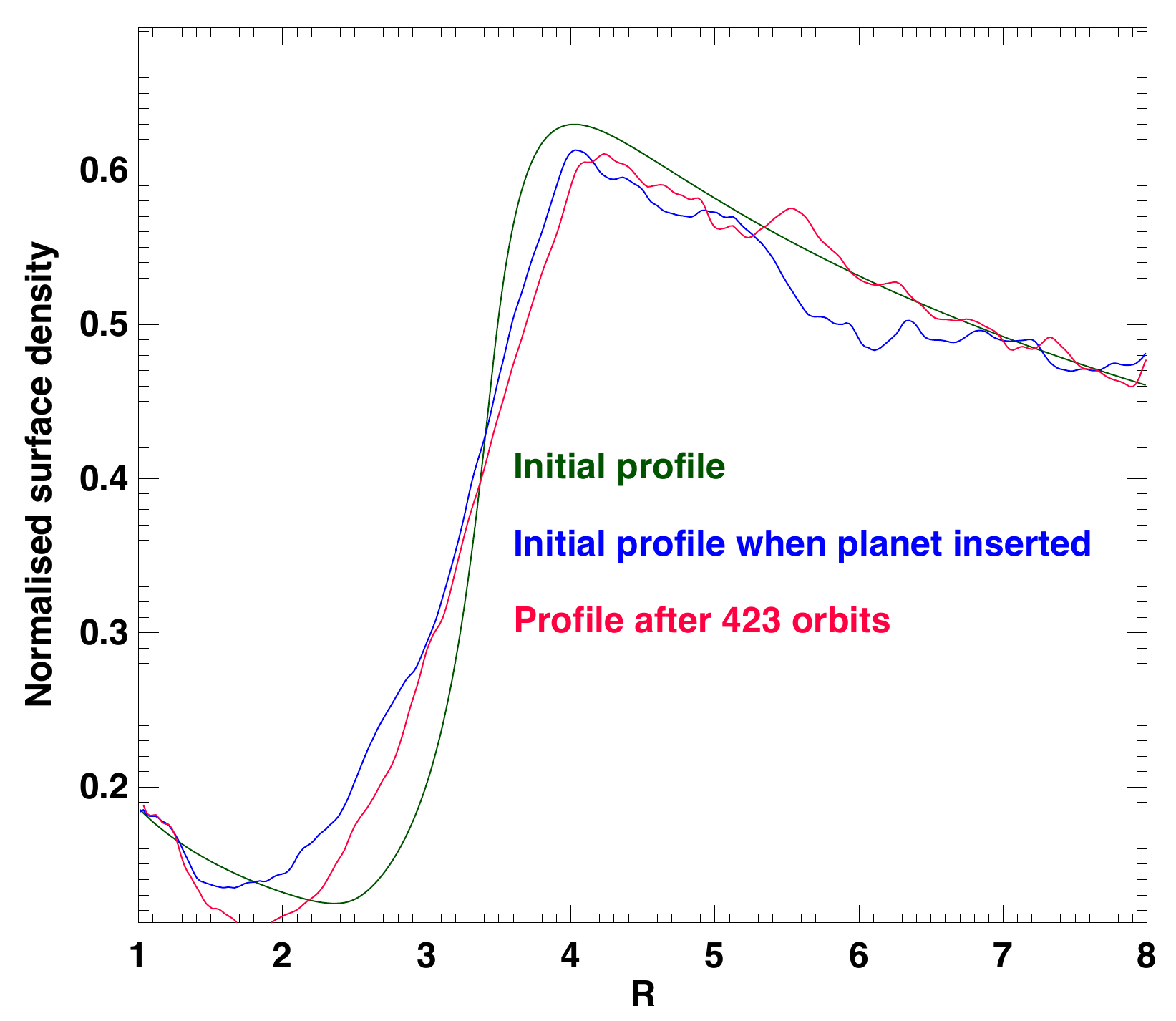}
\end{minipage}
\begin{minipage}{0.49\textwidth}
\includegraphics[width=0.99\textwidth]{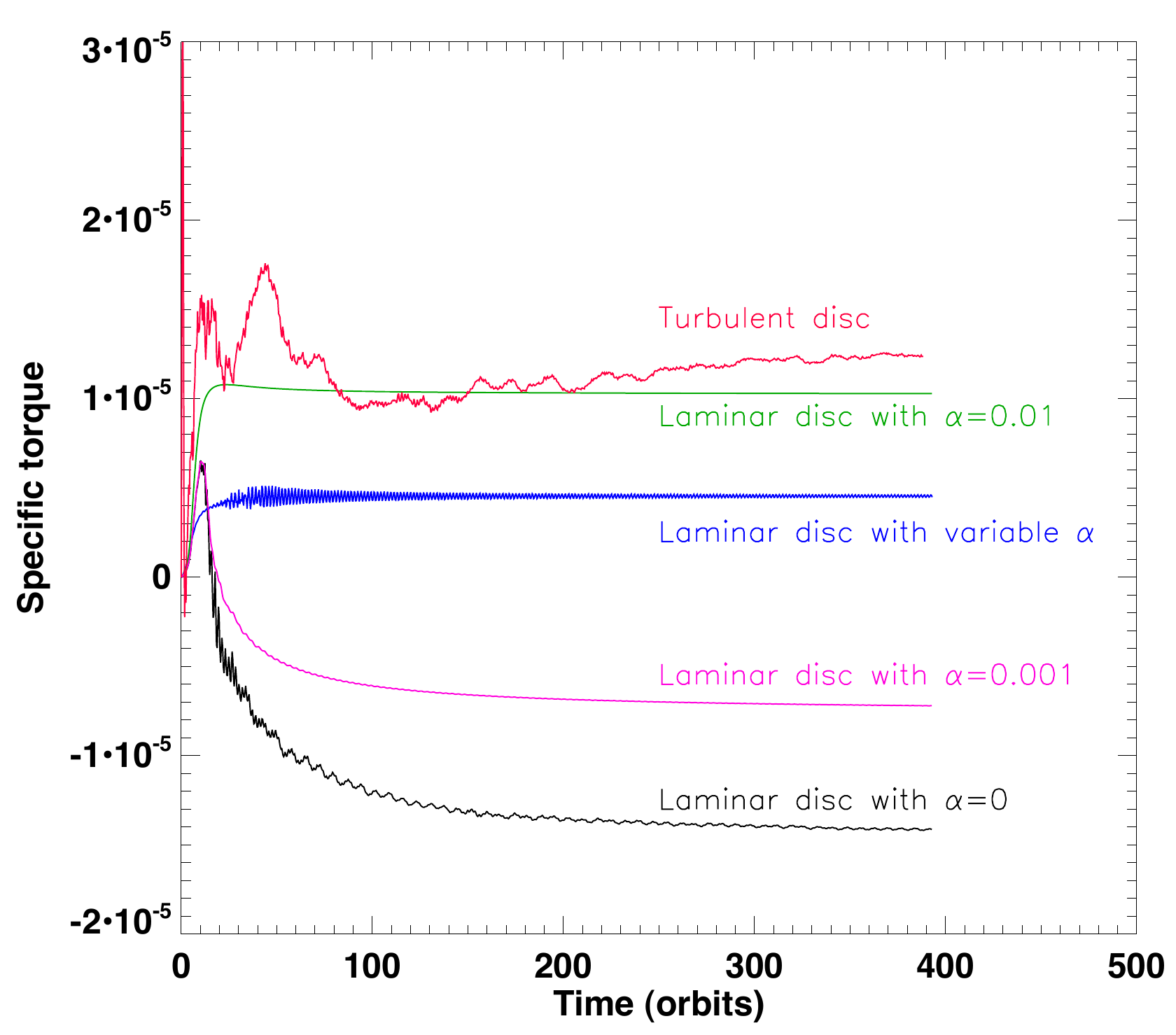}
\end{minipage}
 \caption{{\bf Left:} Surface density profile [normalised by $\Sigma(r_{\rm p})$]
for the simulation with a planet orbiting at the edge of an inner cavity/planet trap
($r_{\rm p}=3$) in a turbulent disk. {\bf Right:} Running averaged torques
experienced by the planet in the turbulent disk, and those experienced by planets
in equivalent laminar models with different values of the viscosity.}
\label{fig:turb-trap}
\end{center}
\end{figure}

Customised MHD simulations designed to specifically examine the
issue of corotation torque saturation in turbulent disks have
been presented by \cite{2011A&A...533A..84B}. Given the long runs times
required for torque convergence, these simulations examined the torque
experienced by Saturn-mass planets in relatively thick disk models
($H/R=0.1$) for which the turbulent stress generated $\alpha \simeq 0.01$-0.03,
and these planets remain in the non gap-forming type~I regime.
Disk models were considered with either power-law surface density profiles,
or with profiles that included a region with a strongly positive
surface density gradient that generates a large corotation torque
(i.e. a planet trap). The surface density profile
from these latter models is shown in the left panel of Fig.~\ref{fig:turb-trap}.
The planet is held on a fixed circular orbit
at radius $R_{\rm p}=3$, and the running time averages of the torques experienced
are shown in the right panel of Fig.~\ref{fig:turb-trap}. 
The average torque for the planet in the turbulent disk maintains a
positive value, indicating a tendency for outward migration.
The value of the torque converges toward the value obtained in a viscous
laminar disk simulation with $\alpha \simeq 0.01$, similar to that obtained
in the MHD simulation, demonstrating clearly that an unsaturated corotation
torque operates. The right panel in Fig.~\ref{fig:turb-trap} also shows
the torques obtained from a number of laminar disk simulations with
different levels of viscosity and shows that, in the absence of viscosity,
the positive corotation torque expected from a `planet trap' fully
saturates, leaving just the negative differential Lindblad torque operating.

The results for power-law disk models from \cite{2011A&A...533A..84B}
were compared with equivalent viscous laminar disk simulations
and with analytic torque estimates, and were found to be in good agreement,
demonstrating further the existence of unsaturated corotation torques
in turbulent disks. 
\cite{2011ApJ...736...85U} presented simulations for a range of planet masses 
embedded in 3D stratified turbulent disks, and for intermediate-mass
planets located in regions where the surface density gradient was
positive locally in their disks, positive corotation torques
were observed. In summary, evidence suggests that turbulent stresses do indeed 
act like viscous stresses and prevent saturation of corotation torques.

A number of unexplored issues remain, however. Do low-mass planets experience 
corotation torques if they can only induce fluid velocities around the
horseshoe U-turns that are smaller than the typical turbulent
velocity ? If so, is the value of the torque experienced equal to that obtained 
from linear theory, as expected in the high-viscosity limit ?
Do the reduced turbulent stresses in a dead zone lead to saturation of corotation 
torques for embedded low-mass planets ?

\subsubsection{Effect of a dead zone}
\label{subsubsec:dead-zone}
There have been numerous numerical studies of the dynamics of protoplanetary disks
with dead zones \citep{2003ApJ...585..908F,2008A&A...483..815I,2008ApJ...679L.131T,
2009ApJ...704.1239O,2011MNRAS.415.3291G,2012MNRAS.420.2419F}.
These indicate that the Reynolds stress at the midplane, induced by sound waves propagating 
into the dead zone from the active layer, has an effective viscosity parameter $\alpha \sim 10^{-4}$.
Assuming that this Reynolds stress acts to prevent saturation of the corotation
torque in the same way as full MHD turbulence, we can estimate
the planet mass for which the horseshoe drag is fully unsaturated. We assume
that this occurs when the viscous diffusion time is approximately equal to half
the horseshoe libration time, such that the corresponding planet-star mass ratio
is
\begin{equation}
\frac{m_{\rm p}}{M_*} \simeq (1.1)^{-2} \left(\frac{4 \pi \alpha}{3} \right)^{2/3} 
\left(\frac{H}{R_{\rm p}} \right)^{7/3},
\label{eq:turb-corot-saturate}
\end{equation}
where we have used eq.~(\ref{eq:kley-nelson-xs}) for the horseshoe width, $x_{\rm s}$,
with $C(\epsilon)=1.1$.
For a dead zone with midplane $\alpha=10^{-4}$ in a disk with typical parameters
($H/R \simeq 0.05$ at 5 AU), a fully unsaturated horseshoe drag occurs for
$m_{\rm p} \simeq  1.4 \; \Me$ (assuming a Solar mass central star).
Planets with masses larger than this are subject
increasingly to saturation of their corotation torques in the manner displayed in
Fig.~\ref{fig:kley-nelson-saturat}, where the total torque experienced by a $20 \; \Me$ planet is displayed as
a function of disk viscosity. This suggests that planets in the super-Earth- and
Neptune-mass range may experience rapid inward migration if embedded in a disk with
parameters similar to those that arise from the above cited simulations
of dead zones.

\subsection{Massive gap opening planets}
\begin{figure}[ht!]
\begin{center}
\begin{minipage}{0.46\textwidth}
\includegraphics[width=0.99\textwidth]{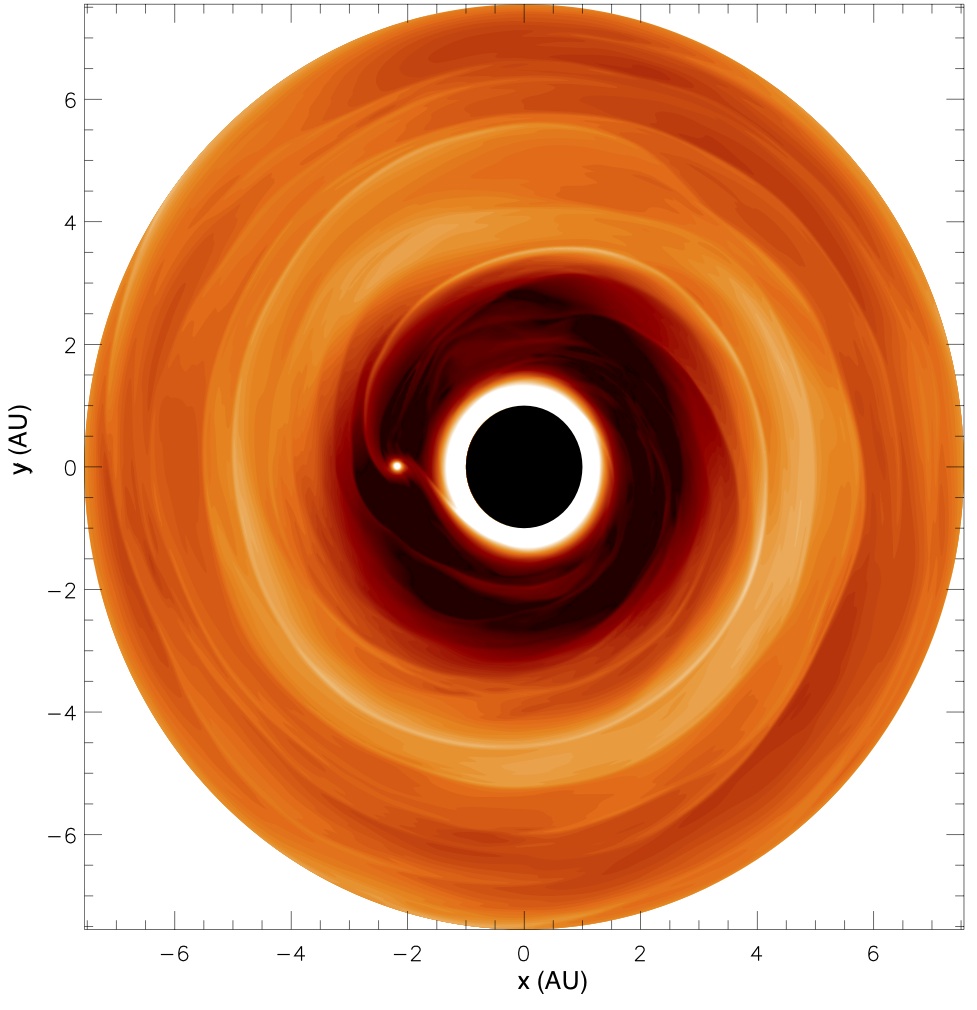}
\end{minipage}
\begin{minipage}{0.51\textwidth}
\includegraphics[width=0.99\textwidth]{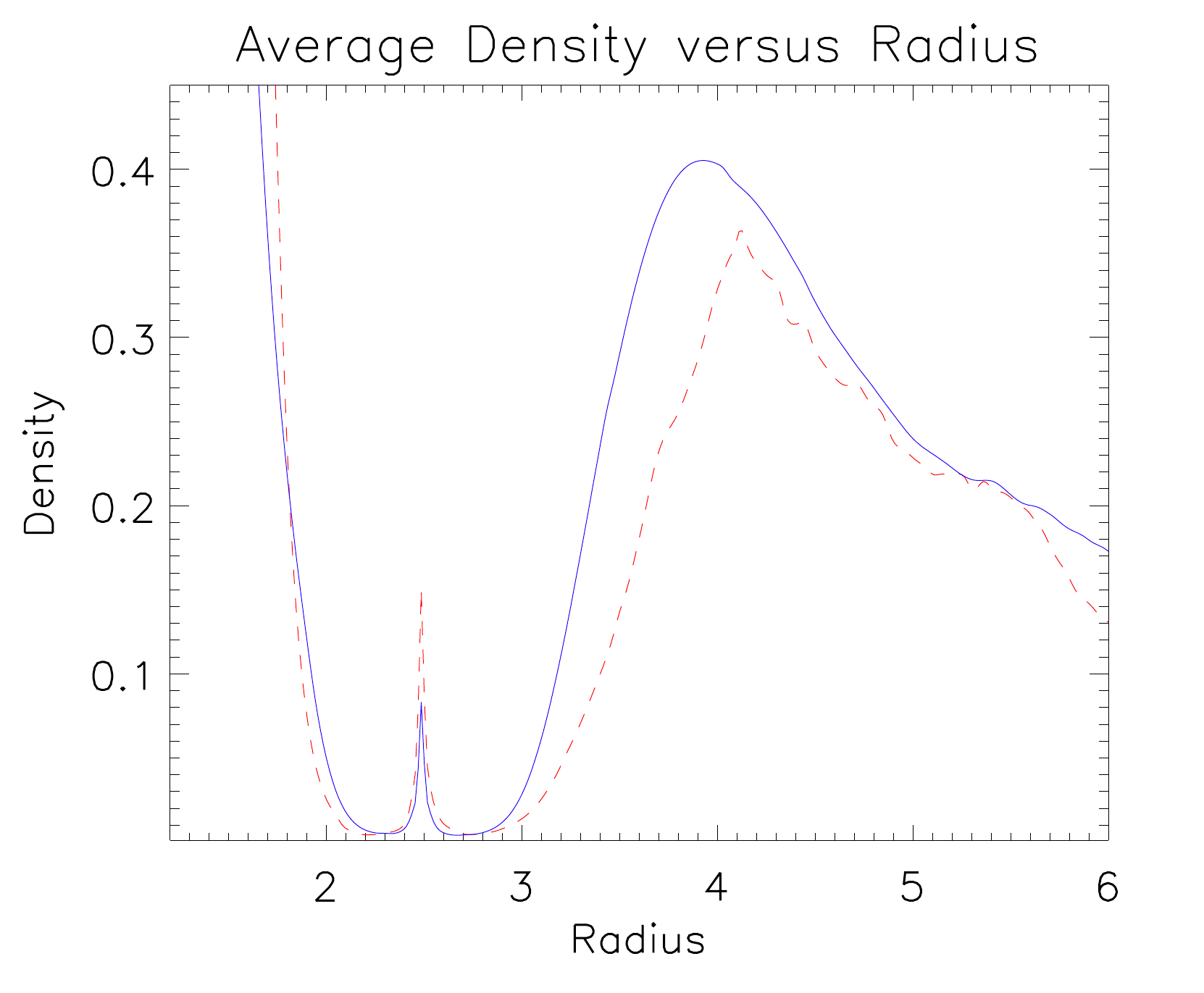}
\end{minipage}
\caption{{\bf Left}: Snapshot of the midplane density for
a 5 Jupiter mass planet orbiting in a turbulent disk with $H/R=0.1$.
{\bf Right:} Normalised midplane density profiles for a magnetised
turbulent disk, and an equivalent laminar disk, containing an
embedded giant planet.}
\label{fig:turb-highmass}
\end{center}
\end{figure}

There has only been a modest amount of research examining
the evolution of giant, gap-forming planets embedded in 
turbulent disks, and to date all studies have adopted the 
ideal MHD approximation. Global, cylindrical models of turbulent 
disks with embedded giant planets have been presented by \cite{2003MNRAS.339..993N},
\cite{2003ApJ...589..543W}, \cite{2004MNRAS.350..849N}, and
\cite{2004MNRAS.350..829P}. Shearing box
simulations were also presented by \cite{2004MNRAS.350..849N}
and \cite{2004MNRAS.350..829P}. More recently, global simulations of vertically
stratified disks have been presented by \cite{2011ApJ...736...85U}. These simulations
all show gap formation for planets with masses in the range of $1 \le m_{\rm p} \le 5$
M$_{\rm Jup}$ for disks with $0.05 \le H/R \le 0.1$ and with effective
stress parameters $\alpha \sim$ few $\times 10^{-3}$, in agreement
with the gap-formation criteria discussed in Sect.~\ref{subsubsec:gap-formation}.
A snapshot from a simulation discussed by \cite{2003MNRAS.339..993N} is displayed in the
left panel of Fig.~\ref{fig:turb-highmass}, showing the formation and
maintenance of a gap by a $m_{\rm p}= 5$ M$_{\rm Jup}$ planet in a disk with
$H/R=0.1$ and $\alpha \simeq 5 \times 10^{-3}$.

Comparisons between the MHD simulations and equivalent laminar, viscous
hydrodynamic simulations show broad agreement in global
properties, but also some interesting differences.
For example, good agreement is found for the migration
timescales of giant planets in turbulent disks, these
being approximately equal to the viscous timescale of the disks
at the planet locations $\simeq 10^5$ yr \citep{2003MNRAS.339..993N,
2004MNRAS.350..849N,2011ApJ...736...85U}. It is generally found that the
gap formed in turbulent disks is wider than that obtained in
equivalent laminar simulations, a point illustrated
by the right panel of Fig.~\ref{fig:turb-highmass} that shows
the azimuthally averaged surface density profile for a laminar and
a turbulent disk simulation discussed by \cite{2004MNRAS.350..829P}.
It is also found that the magnetic field in the turbulent disk simulations
becomes compressed and ordered in the post-spiral shock region near the planet,
and is advected into the Hill sphere of the planet along with the gas.
This latter effect may lead to magnetic braking of the circumplanetary disk, 
leading to enhanced accretion onto planets in turbulent disks with magnetic fields 
compared to non magnetised disks.
\citep{2003MNRAS.339..993N,2004MNRAS.350..829P}. The right panel of
Fig.~\ref{fig:turb-highmass} demonstrates this effect through the enhanced
density of material that has accumulated onto the planet (modeled as
a non-accreting, softened point mass) in the magnetised run. This effect,
however, may be substantially modified when non-ideal MHD effects
such as Ohmic resistivity and ambipolar diffusion are taken into account.
The high surface density circumplanetary disk that forms
around the growing giant planet will probably be largely neutral and not 
strongly coupled to the magnetic field, except for a thin active layer 
at the surface in analogy with dead zones discussed in the previous sections.
Given that the rate of accretion onto a giant planet is determined by
the rate of angular momentum transport within the circumplanetary disk,
this is an important effect for future study as it will play a crucial role
in setting the mass distribution function for fully formed giant planets.

\section{Planets formed through gravitational instability}
Shortly after formation of the central protostar and 
protostellar disk by collapse of a rotating molecular cloud core, 
the disk mass is comparable to the mass of the protostar
and its self-gravity is dynamically important.
The formation of planets by direct fragmentation of self-gravitating
protostellar disks has been the focus of research over a
number of decades \citep{1951PNAS...37....1K,1978M&P....18....5C,1998ApJ...503..923B}.
A general consensus has begun to emerge that fragmentation is most likely to
occur in the outer regions ($R>50$ AU) of protostellar disks where they are
optically thin to their own thermal emission during this early stage.
The gas is able to cool in these regions on timescales less than about the local
orbital period, allowing gravitationally unstable clumps to collapse to
high densities \citep{2001ApJ...553..174G}. The optically thick inner disk, however,
allows only inefficient cooling that prevents fragmentation.
The discovery of massive planets orbiting at large radii from their stars
by direct imaging has led to suggestions that these objects may have formed
via gravitational instability.

Once self-gravitating clumps have formed, it is important to
consider the subsequent long-term evolution of their orbits and masses.
\cite{2005ApJ...633L.137V,2006ApJ...650..956V} presented simulations
of disk and protostar formation through cloud collapse, followed by
disk fragmentation into a succession of protostellar/protoplanetary clumps
that underwent rapid migration through the disk and the inner boundary of
the computational domain (located at a radius of 10 AU). Customised simulations
have been presented more recently \citep{2011MNRAS.416.1971B,2011ApJ...737L..42M}
that examine the migration of planets of different masses in self-gravitating
disks that support gravito-turbulence without fragmenting. In the
gravito-turbulent state, the disk is in near-equilibrium with
heating generated by spiral shocks balancing radiative cooling on average
[although additional heat sources such as stellar radiation may also be important
\cite{2008ApJ...673.1138C,2012ApJ...746..110Z}].
The simulations considered planets with masses ranging
from a Saturn mass up to 5 $M_\mathrm{Jup}$, and showed inward migration 
on type~I migration timescales, albeit with a stochastic component
superposed because of interaction with spiral shocks generated by the disk
self-gravity. 
Planets with 5 $M_\mathrm{Jup}$ masses migrated from 100 AU to
30 AU in $\simeq 3 \times 10^{3} \, {\rm yr}$ 
($\simeq 3$ initial orbital periods) \citep{2011MNRAS.416.1971B},
demonstrating that very rapid inward migration of isolated self-gravitating
clumps is to be expected. Under conditions more appropriate to later stages of
disk evolution, where self-gravity is not important, such a
massive planet would form a gap in the disk and migrate slowly.
The heavier disks that sustain gravito-turbulence considered by
\cite{2011MNRAS.416.1971B}, however, have larger vertical scale heights and
effective viscosities such that a 5 $M_\mathrm{Jup}$ 
protoplanet does not satisfy the gap formation criterion given by
eq.~(\ref{eq:kley-nelson-gap-crida}). 
In the situation of lower mass self-gravitating disks that do not display gravito-turbulence
\citet{2012MNRAS.421..780L} have shown recently that unstable modes
at the gap edges can drive outward type II migration.

A number of uncertainties remain concerning the evolution of
protoplanetary clumps formed by gravitational instability.
An obvious one is understanding the long term orbital
evolution of systems that form multiple clumps
contemporaneously that then interact gravitationally.
\cite{2010Icar..207..509B} discuss one such simulation
in which a clump is scattered onto a highly eccentric
orbit, leading to tidal destruction at periastron.
\cite{2011MNRAS.415.3319C} show examples of fragmenting
disks that result in multiple fragments, some of which are
scattered to large radius and do not migrate inward rapidly.
Perhaps the most significant uncertainty relates to the rate of mass accretion
that can be sustained onto a clump, which needs to lose angular
momentum in order to contract and allow disk material to accrete
into its Hill sphere \citep{2010Icar..207..509B}. The mass
accretion history will clearly affect the migration history,
as rapid mass accretion may allow migration to slow down
through gap formation. The rate of contraction
and growth of the central temperature will also influence the
ability of a clump to survive tidal destruction if it migrates
inward, determine whether or not the growth and sedimentation
of grains can lead to differentiation and core formation
\citep{2011Icar..211..939H}, and consequently determine whether or
not tidal downsizing of inwardly migrating protoplanetary clumps
can lead to the formation of super-Earth planets \citep{2011MNRAS.415.3319C}.
Clearly, substantial further research is required to clarify
the role of gravitational instability and subsequent
disk-protoplanet interactions in the formation of planetary systems.
But it seems to be clear that a single planet-mass object that forms in
a massive self-gravitating disk is very likely to experience rapid
migration into the inner disk.


\begin{figure}[ht!]
\begin{center}
 \includegraphics[width=0.60\textwidth]{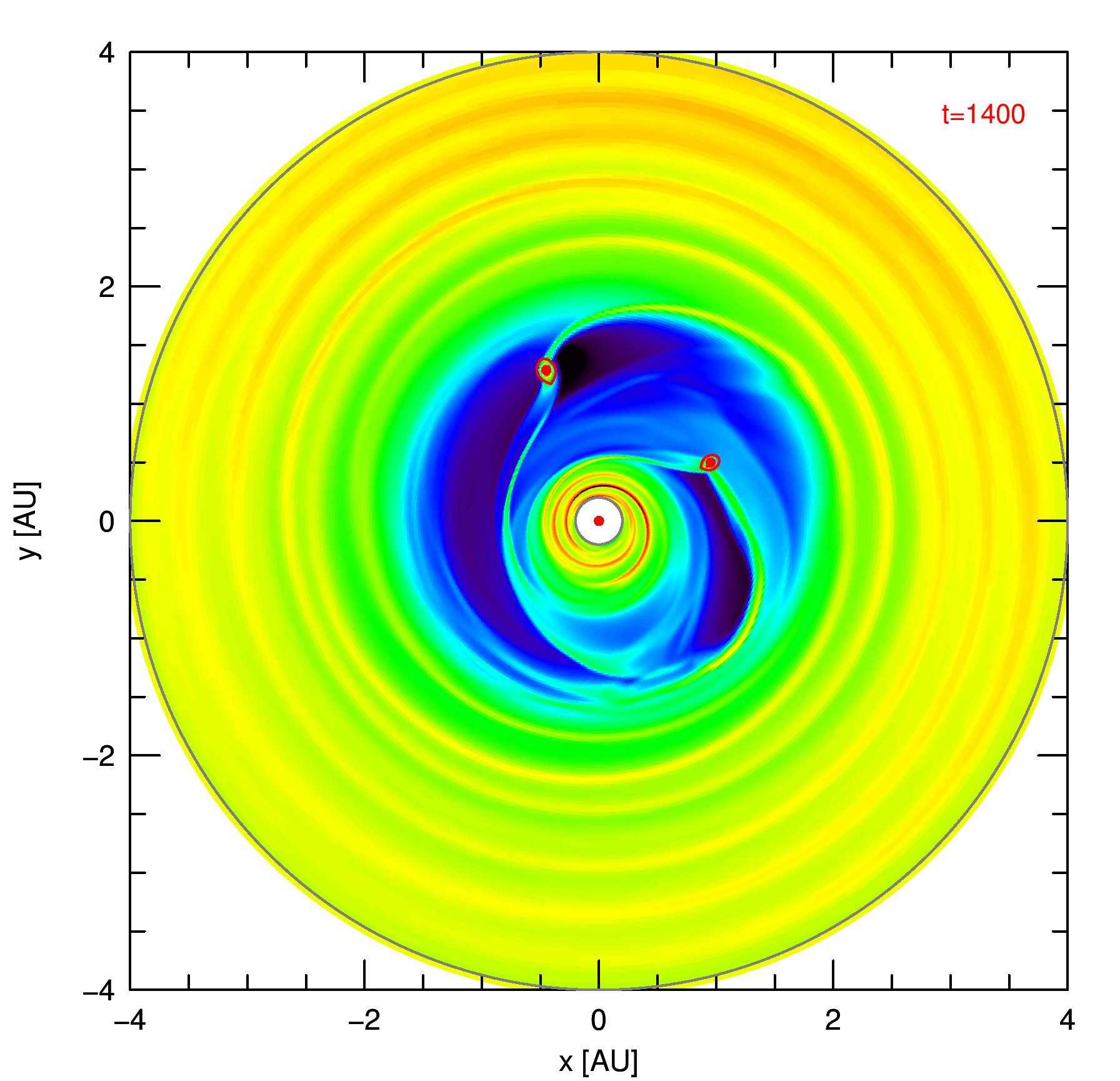}
 \caption{Density structure for two embedded planets engaged in a 2:1 resonance.
    The red lines indicated the Hill radius sizes of the two planets.
    The two planets orbit in a joint wide gap and each of the planets feels
    only a one-sided disk, which helps to maintain the resonant condition.
 }
   \label{fig:kley-nelson-resonance}
\end{center}
\end{figure}

\section{Multiple planets}
\label{sect:multi-planet}

A number of multiple extrasolar planetary systems are known to have at least
two planets in mean motion resonance, and the mere
existence of these systems is strong evidence that dissipative mechanisms
changing the semi-major axes of planets must have operated. The
probability of forming these configurations in situ or later
through mutual scattering is likely to be small [for a different
perspective, however, see  \citet{2008ApJ...687L.107R}].
Convergent migration of a pair of planets will lead
to resonant capture under quite general conditions.
Planets can approach each other from widely separated orbits if they have different migration speeds,
or if they form in close proximity and are sufficiently massive to form a joint
gap (as illustrated in Fig.~\ref{fig:kley-nelson-resonance}).
In this case, the outer disk pushes the outer planet inward,
and the inner disk pushes the inner planet outward, causing convergence.
If the planets approach a commensurability, where the orbital periods
have a ratio of two integers, orbital eccentricities will be excited
and resonant capture may arise. Whether or not this happens hinges primarily 
on the time the planets take to cross the resonance. This must be longer than 
the resonance libration time, which sets a limit on the relative migration speed of 
of the planets \citep{2001A&A...374.1092S}. Resonant capture increases the
eccentricities, and the two planets migrate (usually inward) as a joint pair 
maintaining the commensurability. Continued migration in resonance drives the 
eccentricities upward, and they only remain small during migration if the planets also
experience eccentricity damping \citep{2002ApJ...567..596L}.
If the planets orbit within the disk, as depicted in Fig.~\ref{fig:kley-nelson-resonance}, 
the inner and outer disk act as damping agents, and it has been shown using 2D simulations
that the eccentricity damping rate (due to either the inner or outer disk) for massive planets 
is about 5-10 times faster than the induced migration rate \citep{2004A&A...414..735K}. 
The question of which resonance the system ends up in
depends on the masses, the relative migration speed and initial separation of 
the planets \citep{2002MNRAS.333L..26N}.
Because the 2:1 resonance is the first first-order resonance that two initially
well-separated planets encounter during convergent migration, it is common
for planets to become locked in this commensurability provided migration is not
too rapid.

Application of these results to the best observed system GJ~876 
\citep{2001ApJ...556..296M,2005ApJ...622.1182L}
leads to excellent agreement between theoretical evolution models and observations
\citep{2002ApJ...567..596L,2005A&A...437..727K,2008A&A...483..325C}.
Because formation of resonant planetary systems depends on disk-planet interaction, the
present dynamical conditions in observed systems present an ideal indicator of evolutionary history.
This has been noticed recently in the system HD~45364, where two planets in 
3:2 resonance have been discovered by \citet{2009A&A...496..521C}.
Fits to the data give semi-major axes $a_1 = 0.681$AU and $a_2=0.897$AU, and eccentricities 
$e_1 = 0.168$ and $e_2=0.097$, respectively.
Non-linear hydrodynamic planet-disk models have been constructed for this system
by \citet{2010A&A...510A...4R}. For suitable disk parameters, the planets enter 
the 3:2 resonance through convergent migration.
After the planets reached their observed semi-major axis, a theoretical
RV-curve was calculated. Surprisingly, even though the simulated eccentricities ($e_1 = 0.036, e_2 = 0.017$)
differ significantly from the data fits, the theoretical model fits the observed data points as 
well as the published best-fit solution \citep{2010A&A...510A...4R}.
The pronounced dynamical differences between the two orbital fits, that both match the existing data,
can only be resolved with more observations. Hence, HD~45364 serves as an excellent example
of a system in which a greater quantity and quality of data will constrain theoretical models
of this interacting multi-planet system.

Another interesting observational aspect where convergent migration may have played an important role is the
high mean eccentricity of the observed extrasolar planets. As discussed above, for single planets, the 
disk's action will
nearly always lead to damping of eccentricity (or at best modest growth for Jovian masses).
Strong eccentricity excitation will occur, however,
during convergent migration and resonant capture of two planets.
In the end-phase of the planet formation process the disk will slowly
dissipate and the damping will be strongly reduced. This may leave the resonant planetary system in
an unstable configuration, triggering dynamical instabilities \citep{2003Icar..163..290A}.
Through the subsequent dynamical scattering between the planets, their eccentricities
can be pumped up to higher values.  This scenario has been proposed to explain the
observed wide eccentricity distribution of extrasolar planets 
\citep{2008ApJ...686..580C,2008ApJ...686..603J,2010ApJ...714..194M}.

The simultaneous disk-planet interaction of a set of low-mass embedded planets $(5-20 \; \Me)$ undergoing
differential type~I migration leads to crowded systems as shown by
\citet{2008A&A...482..677C}. These researchers also find
that the planets often form resonant groups with first-order mean-motion resonances having
commensurabilities between 3:2 - 8:7 [see also \citet{2005AJ....130.2884M, 2005MNRAS.363..153P}].
Strong eccentricity damping allows these systems to remain stable during their
migration. In general terms, these systems are reminiscent of the systems of
low-mass planets being discovered by the Kepler mission, such as Kepler-11
\citep{2011Natur.470...53L}.
The proximity of the planets to the star in this system
and their near coplanarity hints strongly toward a disk-driven migration scenario for the
formation of this system.

Another highly interesting dynamical situation occurs for a pair of embedded planets when the outer planet is
less massive and in a regime susceptible to rapid type~III migration. In this case, the outer planet
migrates inward more rapidly than the inner one such that the 2:1 resonance may be crossed, and capture occurs
into the 3:2 resonance \citep{2001MNRAS.320L..55M}. This is, in fact, the preferred outcome if the outer planet has about
a third of the mass of the inner planet \citep{2008A&A...482..333P}.
Suprisingly, after capture, both planets begin to migrate {\it outward},
maintaining the 3:2 resonance \citep{2001MNRAS.320L..55M}.
This occurs because the two planets form a joint gap, but one that has
an asymmetry between its inner and outer edges because of the
lower-mass of the outer planet. This situation allows the inner
disk to exert stronger torques than the outer disk, leading
to outward migration. The situation is assisted by the fact that the lower mass
outer planet funnels mass through the gap and eventually into the inner disk,
thus maintaining a high surface density there.
One application of the Masset-Snellgrove process refers to the formation of systems 
such as HR~8799 or Fomalhaut where the planets reside at very large distances from the 
star \citep{2008Sci...322.1348M,2008Sci...322.1345K}.
Here, it has been suggested that this very mechanism can take the planets, under suitable disk conditions,
all the way from about 10~AU up to 100~AU \citep{2009ApJ...705L.148C}.

As explained above and shown in Fig.~\ref{fig:kley-nelson-resonance}, multiple planets carve very wide gaps in the
disk. The situation where the inner disk has been completely cleared, for example accreted onto the star,
gives rise to so-called transitional disks that display a large inner hole, inferred by the lack of near-IR
radiation in their SEDs. This possibility has been studied by \citet{2004ApJ...612L.137Q},
\citet{2006ApJ...640.1110V}, \citet{2011ApJ...738..131D}, and \citet{2011ApJ...729...47Z}.

\section{Planet formation with migration: comparison with observations}
\label{sect:migrat-obs}
The migration scenarios discussed in this review depend
on the planet mass, so it is crucial to consider the combined effects
of mass growth and migration when assessing the role of migration 
during the formation of planetary systems.
Two approaches that have been used extensively for this purpose are
planetary population synthesis and N-body simulations.
Population synthesis studies use Monte-Carlo techniques to construct synthetic
planetary populations for comparison with exoplanet observations, with the aim of determining
which combinations of model ingredients lead to statistically reasonable fits to the observational
data (orbital elements and masses in particular).
The basis of published models has been the core-accretion scenario of planetary formation,
combined with simple prescriptions for type~I and type~II migration and viscous disk evolution models
\citep{2002MNRAS.334..248A,2004ApJ...604..388I,2005A&A...434..343A}.
Almost all studies to date have adopted type~I migration rates according to 
eq.~(\ref{eq:kley-nelson-gammatot}),
supplemented with a reduction factor that slows type~I migration.
The influence of the vortensity and entropy-related horseshoe drag discussed in Sect.~\ref{subsec:adi-rad}
has not yet been explored in general, although there are a couple of recent exceptions to this statement.

\cite{2008ApJ...673..487I}, \cite{2009A&A...501.1139M}, and
\cite{2009A&A...501.1161M} consider the effects of type~I and type~II migration in their
population synthesis models. Although differences exist in the modeling procedures, 
these studies conclude that
unattenuated type~I migration leads to planet populations
that do not match the observed distributions of planet mass and
semimajor axis. Models presented by \cite{2008ApJ...673..487I}
fail to produce giant planets at all if full-strength type~I migration
operates. Acceptable planet populations are reported for reductions
in the efficiency of type~I migration by factors of 0.01 to 0.03,
with type~II migration being required to form `hot Jupiters'. 
With the type~II timescale of $\sim 10^5$ yr being
significantly shorter than disk life-times, numerous
giant planets also migrate into the central star in these models.
The survivors are planets that form late as the disk
is being dispersed (through viscous evolution and photoevaporation).
\cite{2009A&A...501.1139M} and \cite{2009A&A...501.1161M} present models
with full-strength type~I migration that are able to form a sparse
population of gas giants.
Cores that accrete late in the disk life-time are
able to grow to large masses as they migrate because they
do not exhaust their feeding zones. Type~I migration in this case
leads to too many short-period massive gas giants
that contradict the exoplanet data. 

The above studies focused primarily on forming gas giant planets,
but numerous super-Earth- and Neptune-mass planets have been
discovered. In a recent study, \cite{2010Sci...330..653H}
compared the predictions of population synthesis models
with radial-velocity observations of extrasolar planets
orbiting within 0.25 AU around 166 nearby G-, K-, and M-type
stars (the $\eta_{\rm Earth}$ survey). The data indicate
a high density of planets with $m_{\rm p}=$ 4 - 10 M$_{\oplus}$ 
with periods of less than 10 days, which is not present in the population 
synthesis models because of rapid migration and mass growth.
\cite{2010ApJ...719..810I} recently considered the formation
of super-Earths using population synthesis.
In the absence of an inner disk cavity (formed by interaction with
the stellar magnetic field) the simulations failed to form 
systems of short-period super-Earths because of type~I migration.

N-body simulations with prescriptions for migration have
also been used to examine the interplay between
planet growth and migration. This approach has the advantage of
including important planet-planet interactions that can induce
orbital changes, as opposed to the `single-planet in a disk' approach
adopted by most population synthesis studies.
Simulations that explore short-period super-Earth formation and
tidal interaction with the central star for disk models
containing inner cavities  have been presented by
\cite{2007ApJ...654.1110T}.
\cite{2009MNRAS.392..537M,2010MNRAS.401.1691M}
considered the formation of hot super-Earth- and Neptune-mass
planets using N-body simulations combined with
type~I migration. They examined whether or not the standard
oligarchic growth picture of planet formation combined with
migration could produce systems such as Gliese 581 and
HD 69830 that contain multiple short-period super-Earth- and Neptune-mass planets.
These super-Earth and Neptune systems probably
contain up to 30 -- 40 Earth masses of
rocky or icy material orbiting within 1 AU, but the simulations failed
to produce any systems with greater than 12 Earth masses of solids
interior to this orbital radius.

\cite{2008Sci...321..814T} presented a 
suite of simulations of giant planet formation using a hybrid
code in which emerging embryos were evolved using an N-body
integrator combined with a 1D viscous disk model. Although
unattenuated type~I and type~II migration were included, a number of
models led to successful formation of systems of
surviving gas giant planets. These models consider an initial
population of planetary embyos undergoing oligarchic growth extending
out to 30 AU from the star, and indicate that the right combination of
planetary growth times, disk masses and life times can form 
surviving giant planets through the core-accretion model provided
embryos can form and grow at large orbital distances before migrating
inward.

The role of the combined vorticity- and entropy-related corotation torque,
and its ability to slow or reverse type~I migration
of forming planets, has not yet been explored in detail.
The survival of protoplanets with masses in the
range of $1 \le m_{\rm p} \le 10 \; \Me$ in global 1D disk models
has been studied by \citet{2010ApJ...715L..68L}. Their
models demonstrate that there are locations in the disk
where planets of a given mass experience zero migration
due to the cancellation of Lindblad and corotation torques
(zero-torque radii).
Planets have a tendency to migrate toward these positions,
where they then sit and drift inward slowly as the gas disk 
disperses. 
Preliminary results of population synthesis calculations
have been presented by \cite{2011IAUS..276...72M},
and N-body simulations that examine the oligarchic
growth scenario under the influence of strong corotation
torques have been presented by \cite{2012MNRAS.419.2737H}
These studies indicate that the convergent migration 
that arises as planets move toward their zero-migration
radii can allow a substantial increase in the rate of 
planetary accretion, and the formation of gas giants at 
large distance from the central star. \cite{2012MNRAS.419.2737H}
find, however, that planet-planet scattering can increase the 
eccentricity to values that effectively quench the horseshoe drag, such
that crowded planetary systems during the formation epoch
may continue to experience rapid inward migration. Further
work is clearly required to fully assess the influence
of the corotation torque on planet formation in the prescence of
significant planet-planet interactions.

\subsection{Observational evidence for disk-driven migration}
\label{subsec:migrat-evidence}
The discovery of the first extrasolar planet orbiting around
a solar-type star, 51 Peg \citep{1995Natur.378..355M}, 
provided immediate evidence for the migration of planets. In situ 
formation models cannot explain Jupiter-like planets orbiting 
with periods of a few days because the dust required to build 
a core sublimates at the disk temperatures expected at these
small radii ($T > 1,500$ K).The existence of hot Jupiters, however,
does not tell us which of the various modes of migration that
we have discussed are dominant during or after 
planet formation. In principle, type~I migration of a 
solid core followed by in situ accretion of gas may explain
hot Jupiter systems, as can a picture based on near-complete 
formation at large orbital radii followed by type~II migration inward.

Recent observations of the {Rossiter-McLaughlin effect}
indicate the presence of short-period planets whose
orbit planes are strongly misaligned with their host star equatorial planes
\citep{2010A&A...524A..25T,2010ApJ...718L.145W}. This dynamical feature
may be difficult to reconcile with disk-driven migration that comes about
with inclination damping. As an alternative, it has been suggested that different processes
such as planet-planet scattering \citep{2008ApJ...686..580C}
and/or the {Kozai mechanism} combined with stellar tides 
\citep{2007ApJ...669.1298F} may play an important role.
The most natural explanation for short-period systems whose orbital angular momenta are aligned
with the stellar spin axis, however, remains to be disk-driven migration. 
Furthermore, planets with intermediate periods of a few tens
of days cannot have their migration explained by scattering or
the Kozai mechanism as stellar tides do not operate effectively 
in these systems.

The discovery of multiple systems of super-Earth and Neptune-like 
planets orbiting interior to $\sim 0.5$ AU, such as Gliese 581
\citep{2007A&A...469L..43U} and HD 69830 \citep{2006Natur.441..305L} 
seem to provide compelling evidence for type~I
migration of low-mass planets, albeit at slower rates than
the standard values. In situ formation is very difficult
to explain for these systems, which likely contain up to 30 - 40 Earth
masses of rocky and icy material. The mass of solids
interior to 1 AU in typical disk models is at most a few Earth masses.
The discovery by the Kepler mission of multiple systems of low-mass planets on
near coplanar orbits in resonant or near-resonant configurations,
such as the Kepler-11 system \citep{2011Natur.470...53L}
provides further compelling evidence for formation and
migration in a highly dissipative disk environment given the apparent
dynamical quiesence of these closely packed systems.

As discussed in Sect.~\ref{sect:multi-planet}, the existence of giant planets
in 2:1 mean motion resonances, such as the GJ~876 \citep{2001ApJ...556..296M}
and Kepler-9 \citep{2010Sci...330...51H} systems,
suggests that the differential migration in these systems
was relatively slow when the resonances were established. The more
rapid migration rates associated with type~I or type~III migration
do not result in capture into 2:1, but in resonances of higher degree
such as 3:2 \citep{2001MNRAS.320L..55M}. Although this evidence is
circumstantial rather than direct, it points to migration rates that
are characteristic of type~II migration. The success of models
based on type~II migration in explaining the basic orbital parameters of 
the GJ 876 system makes this circumstantial evidence even more compelling.

\section{Summary Points}
\label{sect:summary}
\begin{enumerate}
\item Disk-planet interaction is a natural process operating when young
      planets are still embedded in the protoplanetary disk. Torques are created
      by the induced non-axisymmetric features in the disk: spiral waves and the horseshoe region.
      Disk-planet interaction impacts all orbital elements of a planet. Eccentricity and inclination are
      usually damped quickly, but the speed and direction of migration depends intricately on disk physics. 
\item Disk-planet interaction is intrinsically three-dimensional.
      Using the latest numerical methods, new unexpected results, such as outward 
      migration, have been discovered.
\item Standard migration (in viscous, laminar disks) comes in 3 flavours: 
      type~I migration operates for nongap-forming low-mass planets 
      (it can be inward or outward). In type~II migration, the planet opens a gap and moves with the 
      viscously evolving disk. It is typically inward and slower than type~I.
      Type~III describes fast runaway migration driven by the radial drift of the planet
      (it can be either inward or outward).
\item In turbulent, magnetically driven disks, low-mass planets experience 
      stochastic motion.
      The evolution of more massive planets can be described approximately 
      by viscous disk-planet interaction.
\item The disk's thermal structure plays a decisive role in determining the direction of 
      type~I migration.
      For radiative disks, outward migration is possible for planets below about 
      $30 \; \Me$ on circular orbits.
      This effect helps to solve the too-rapid migration problem. 
\item Disk-planet interaction is one major ingredient in shaping the final architecture 
      of planetary systems as we observe them today.
\item Evidence for migration comes from three basic observational facts: 
      {\it i)} Hot Jupiters close to the star with orbital angular momentum 
               vectors approximately aligned to the spin-axis of the star.
      {\it ii)} Coplanar multiple systems containing low-mass planets with short and 
                intermediate orbital periods, as discovered, for example, by the Kepler mission, and
      {\it iii)} The high fraction of mean motion resonances in multi-planet systems.
\end{enumerate}

\section{Future Issues}
\label{sect:future}
\begin{enumerate}
\item Planetary growth and evolution takes place in protoplanetary disks,
      and the final state of a planetary system is determined by the disk's structure.
      Theoretical, computational, and observational developments are required to 
      improve models of the planet formation environment, with particular
      emphasis on their structural evolution over time. Important ingredients are the disk's
      self-gravity, irradiation from the central star, chemistry, and non-ideal MHD processes.
\item Because the planet mass is the important parameter determining the strength of
      planet-disk interaction, its mass growth needs to be known simultaneously 
      with its orbital evolution. Orbital eccentricity also plays a role in
      determining disk torques. Detailed models that combine the 
      temporal history of mass growth, migration, and planet-planet interactions
      are required.
\item Refined torque formulae to be used in population synthesis models are needed
      that take into account the results of full 3D disk-planet modeling. 
\item Improvements in population synthesis-style modeling are required to
      increase the connection between theory and observations. The success of
      missions such as CoRoT and Kepler, continuing discoveries from ground-based
      observatories, and the prospect of future missions such as PLATO or TESS
      make this a high priority.
\item Future ground-based observatories such as ALMA have the potential to observe
      planet-forming disks around young stars in unprecedented detail.
      Improved modeling of processes such as gap formation in more
      realistic disk models will be required to interpret
      future observations.
\end{enumerate}

\section*{Discolosure statement}
The authors are not aware of any affiliations, memberships, funding, or financial holdings
that might be perceived as affecting the objectivity of this review.

\section*{Acknowledgement}
W. Kley acknowledges the support by the German Research Foundation (DFG) through grants within the 
Collaborative Research Group FOR 759:  {\it The formation of Planets: The Critical First Growth Phase}.

%
\bibliography{kna}

\end{document}